\documentclass{aa}  

\usepackage{amsmath}
\usepackage{amssymb}
\usepackage{bm}
\usepackage{cancel}
\usepackage{commath}
\usepackage[thinc]{esdiff}
\usepackage{mathtools}
\usepackage{mathabx}
\usepackage{nicefrac}
\usepackage{relsize}

\usepackage{booktabs}
\usepackage{multirow}
\usepackage{tabularx}

\usepackage{graphicx}
\usepackage{svg}
\usepackage{xcolor}

\usepackage{natbib}
\usepackage[colorlinks=true, allcolors=blue, breaklinks=true]{hyperref}
\usepackage[noabbrev, capitalise]{cleveref}

\usepackage{afterpage}
\usepackage{csquotes}
\usepackage{enumitem}
\usepackage[justification=centering]{subcaption}
\usepackage{xspace}
\bibpunct{(}{)}{;}{a}{}{,} 

\makeatletter
\renewcommand*\aa@pageof{, page \thepage{} of \pageref*{LastPage}}
\makeatother

\newcommand{\gallifrey}{\texttt{gallifrey}\xspace}
\newcolumntype{L}{>{\raggedright\arraybackslash}X}
\newcolumntype{R}{>{\raggedleft\arraybackslash}X}
\newcolumntype{C}{>{\centering\arraybackslash}X}

\begin{document}

   \title{\gallifrey: JAX-based Gaussian process structure learning for astronomical time series}
\titlerunning{\gallifrey: JAX-based GP structure learning for astronomical time series}   

   \author{C. Boettner       \inst{1}
          \fnmsep
          \inst{2}
          }

   \institute{Kapteyn Astronomical Institute, University of Groningen,
              Landleven 12 (Kapteynborg, 5419) 9747 AD Groningen\\
              \email{boettner@astro.rug.nl}
              \and
              GELIFES Institute, University of Groningen,
              Nijenborgh 7 9747 AG Groningen
             }

   \date{}
   
   \abstract
    {Gaussian processes (GPs) have become a common tool in astronomy for analysing time series data, particularly in exoplanet science and stellar astrophysics. However, choosing the appropriate covariance structure for a GP model remains a challenge in many situations, limiting model flexibility and performance.}
    {This work provides an introduction to recent advances in GP structure learning methods, which enable the automated discovery of optimal GP kernels directly from the data, with the aim of making these methods more accessible to the astronomical community.}
    {We present \gallifrey, a JAX-based Python package that implements a sequential Monte Carlo algorithm for Bayesian kernel structure learning. This approach defines a prior distribution over kernel structures and hyperparameters, and efficiently samples the GP posterior distribution using a novel involutive Markov chain Monte Carlo procedure.}
    {We applied \gallifrey to common astronomical time series tasks, including stellar variability modelling, exoplanet transit modelling, and transmission spectroscopy. We show that this methodology can accurately interpolate and extrapolate stellar variability, recover transit parameters with robust uncertainties, and derive transmission spectra by effectively separating the background from the transit signal. When compared with traditional fixed-kernel approaches, we show that structure learning has advantages in terms of accuracy and uncertainty estimation.}
    {Structure learning can enhance the performance of GP 
    regression for astronomical time series modelling. We discuss a road map for
    algorithmic improvements in terms of scalability to larger datasets, so that the methods presented here can be applied to future stellar and exoplanet 
    missions such as PLATO.}
   

   \maketitle
%

\section{Introduction}
\label{sec:introduction}

Exoplanet science and stellar astrophysics are increasingly reliant on the analysis of high-cadence, high-precision time series data from dedicated surveys. Missions like \textit{Kepler} \citep{Borucki2010} and TESS \citep{Ricker2015}, and the forthcoming PLATO mission \citep{Rauer2014b}, generate large quantities of data, requiring robust statistical methodologies for signal extraction and noise characterisation.  Gaussian processes (GPs) have, in this context, emerged as a powerful and versatile tool for a wide variety of applications, including transit searches \citep{Crossfield2016}, transit modelling \citep{Gibson2012, Gibson2014, Barros2020}, systematics modelling \citep{Foreman-Mackey2015, Aigrain2015, Aigrain2016}, low-resolution space- and ground-based spectroscopy \citep{Evans2013, Evans2015, Evans2017, Ahrer2022, Ahrer2023}, stellar variability modelling \citep{Angus2018, Gillen2020, Luger2021, Nicholson2022}, and radial velocity modelling \citep{Aigrain2012, Haywood2014}. GPs are particularly useful in situations involving correlated noise, where neglecting the correlation can lead to an overestimation of the signal-to-noise ratio and, in turn, to false positive detections \citep{Pont2006}. Thanks to this wide applicability, a variety of software packages have been created for GP modelling, including \texttt{george} \citep{2015ITPAM..38..252A}, \texttt{tinygp} \citep{Foreman-Mackey2024}, and \texttt{celerite} \citep{Foreman-Mackey2017a}, as well as packages that directly incorporate GPs into astronomical modelling, such as \texttt{juliet} and \citep{Espinoza2019} \texttt{exoplanet} \citep{exoplanet:exoplanet}. For a comprehensive review of GP use in astronomy, see \citet{Aigrain2023}. 

A key practical challenge in applying GPs, however, lies in selecting an appropriate kernel function capable of adequately representing the data. Traditional approaches, relying on manual selection from a limited set of kernels and subsequent hyperparameter optimisation, can be subjective and may not fully exploit the potential of GP methods for complex time series modelling. With recent advances in machine learning, there has been significant interest in learning the appropriate covariance structure directly from the data instead.

A fruitful approach is to define a flexible symbolic language over possible kernel structures and combinations, and then to search this space for appropriate kernel structures \citep{Duvenaud2013, Abdessalem2017}. In one of the early works in this direction, \citet{Duvenaud2013} employed a greedy-search algorithm to find the best-fitting kernel structure in an iterative fashion. This was followed by \citet{Saad2019}, who introduced a fully Bayesian framework by defining a prior over possible kernel structures and sampling them using Markov chain Monte Carlo (MCMC) methods.  This, in turn, was improved upon by \citet{Saad2023}, who refined the approach of \citet{Saad2019} by introducing a novel sequential Monte Carlo (SMC) algorithm \citep{DelMoral2006, Chopin2020}, enhancing both performance and speed compared to the pure MCMC version.

The main goal of this work is to make these advances accessible to the exoplanet community. To this end, we introduce \gallifrey, a Python package specifically designed to facilitate GP time series structure learning.  \gallifrey implements an SMC algorithm based on \citet{Saad2023}, incorporating a Bayesian prior over kernel structures and leveraging the computational efficiency of the JAX framework \citep{jax2018github} for both enhanced performance and automatic differentiation capabilities. To facilitate the usage of this framework, we provide the open-source code\footnote{\url{https://github.com/ChrisBoettner/gallifrey}} as well as in-depth documentation, which provides detailed tutorials to create all the figures presented in this work.\footnote{\url{https://chrisboettner.github.io/gallifrey}}

The paper is structured as follows: In Sect. \ref{sec:methods} we provide a short review of GPs and GP regression, introduce the theoretical framework for structure learning, and detail the SMC algorithm for efficient sampling of the GP posterior.  In Sect. \ref{sec:gallifrey} we introduce the \gallifrey Python package and discuss its design choices. In Sect. \ref{sec:examples} we demonstrate applications of this methodology to a range of stellar variability, transit modelling, and transmission spectroscopy cases.  Finally, in Sect. \ref{sec:discussion} we discuss use cases of the package and outline a road map for future improvements.

\section{Methods}
\label{sec:methods}
\subsection{A brief review of Gaussian processes}
\label{sec:gpreview}
Gaussian processes are a versatile tool in parameter-independent statistical modelling, and can be understood as a generalisation of the Gaussian probability distribution. By defining a distribution over functions, rather than individual variables, they can be thought of as an infinite-dimensional analogue of a Gaussian distribution. In the following, we give a brief summary of the key equations for GPs and GP  regression. This description closely follows the seminal textbook by \citet{Rasmussen2006a}.
\subsubsection{Definition}
\label{sec:gpdefinition}
A one-dimensional, zero-mean, continuous-time GP is defined through a covariance function $k(x, x')$ as
\begin{equation}
    f(x) \sim \mathcal{GP}(0, k(x, x')),
    \label{eq:gpdefinition}
\end{equation}
where the covariance function is defined as $k(x, x') = \mathbb{E}[(f(x)-m(x))(f(x')-m(x'))]$.\footnote{One can also specify a mean function $m(x) = \mathbb{E}[f(x)]$, but in many applications, the mean function is assumed to be zero for simplicity.} A GP is characterised by the fact that for any finite set of sample points $\bm{x} = [x_1, x_2, \ldots, x_n]$, the probability distribution of the corresponding function values $\bm{f}$ is given by a multivariate Gaussian:
\begin{equation}
    \bm{f} (\bm{x}) \sim \mathcal{N}\left( \bm{0} , K(\bm{x},\bm{x})\right),
    \label{eq:gpsampledistribution}
\end{equation}
where $K$ is the covariance matrix constructed by evaluating the kernel function $k$ at points $\bm{x}$.  For a given kernel, a GP spans a space of functions, with associated probabilities for these functions under the GP. In a regression task, the GP can be treated as a prior over these functions, rather than parameters, and can be used for non-parametric inference.

\subsubsection{Gaussian process regression}
\label{sec:gpregression}
In Bayesian regression of parametric models, priors are typically placed over parameters $\bm{\theta}$ (e.g. $a$ and $b$ in $a \cdot x + b \cdot x^2$), and a likelihood function $\mathcal{L}$ represents the probability of observing data $\mathcal{D} = \{\bm{x},\bm{y}\} = \{x_i, y_i\}_{i=1}^N$ given the model and specific parameter values.  The posterior distribution of these parameters is then derived using Bayes's theorem,
\begin{equation}
    P(\bm{\theta} | \mathcal{D}) ~\propto~ \mathcal{L}(\mathcal{D} | \bm{\theta}) \cdot P(\bm{\theta}),
\end{equation}
where $P(\bm{\theta})$ is the prior over the parameters $\bm{\theta}$. In contrast, GP regression can be understood as defining the prior directly over functions $f$. Given a likelihood function and data, the posterior distribution over functions is given by
\begin{equation}
    P(f | \mathcal{D}) ~\propto~ \mathcal{L}(\mathcal{D} | f) \cdot P(f),
\end{equation}
where the prior $P(f)$ is defined by the GP given by Eq. \ref{eq:gpdefinition}.  As in the parametric case, the choice of likelihood function is flexible and should be selected to appropriately model the noise in the data. In particular, it is not restricted to a Gaussian likelihood, despite the prior being a GP. However, a number of key relations become analytically tractable when a Gaussian likelihood is assumed.

 For a Gaussian likelihood with variance $\sigma_\mathrm{n}^2$, the joint distribution of observed values $\bm{y}$ at $\bm{x}$, and unobserved latent values $\bm{f}^*$ at new points $\bm{x}^*$ is given by
 \begin{equation}
     \left[\begin{array}{l}\bm{y} \\ \bm{f}^*\end{array}\right] \sim \mathcal{N}\left(\bm{0},\left[\begin{array}{cc}K(\bm{x}, \bm{x})+\sigma_\mathrm{n}^2 \bm{I} & K(\bm{x}, \bm{x}^*) \\ K(\bm{x}^*, \bm{x}) & K(\bm{x}^*, \bm{x}^*)\end{array}\right]\right).
 \end{equation}
 
 By marginalising over $\bm{y}$, we obtain the predictive distribution for the latent function values $\bm{f}^*$ at the locations $\bm{x}^*$, which is also given by a Gaussian,
 \begin{equation}
    P\left(\bm{f}^* | \bm{x}^*, \bm{x}, \bm{y}\right) = \mathcal{N}(\bm{\mu}, \bm{\Sigma}),
    \label{eq:gplatentpredictivedistribution}
\end{equation}
where
\begin{equation}
    \bm{\mu} =  K\left(\bm{x}^*, \bm{x}\right)\left[K(\bm{x}, \bm{x})+\sigma_\mathrm{n}^2 \bm{I}\right]^{-1} \bm{y},
\end{equation}
 and
 \begin{equation}
     \bm{\Sigma} = K\left(\bm{x}^*, \bm{x}^*\right)-K\left(\bm{x}^*, \bm{x}\right)\left[K(\bm{x}, \bm{x})+\sigma_\mathrm{n}^2 \bm{I}\right]^{-1} K\left(\bm{x}, \bm{x}^*\right).
 \end{equation}
In most applications, $\sigma_n^2$ will correspond to the known measurement uncertainties of the observations, although it can also be treated as a learnable parameter. The predictive distribution for new observations $\bm{y}^*$ at locations $\bm{x}^*$ is given by
 \begin{equation}
    P\left(\bm{y}^* | \bm{x}^*, \bm{x}, \bm{y}\right) = \mathcal{N}(\bm{\mu}, \bm{\Sigma} +\sigma_\mathrm{n}^2 \bm{I}).
    \label{eq:gppredictivedistribution}
\end{equation}

\newpage
For a set of residuals $\bm{r} = \left(\bm{y}^* - \bm{\mu}\right)$, and a covariance matrix $\bm{\tilde{\Sigma}} = \bm{\Sigma} +\sigma_\mathrm{n}^2 \bm{I}$, we can define the whitened residuals
\begin{equation}
   \bm{z} = \bm{U} \bm{r}, 
   \label{eq:gpwhitenedresiduals}
\end{equation}
where $\bm{U}$ is defined by the Cholesky decomposition $\bm{\tilde{\Sigma}}^{-1} = \bm{U}^\top \bm{U}$. From Eq. \ref{eq:gppredictivedistribution}, follows that the distribution of the whitened residuals $\bm{z}$ is given by the standard Gaussian $\mathcal{N}(0, \bm{I})$. Furthermore, in the case of a Gaussian likelihood, the marginal likelihood (i.e. Bayesian model evidence) $P(\mathcal{D}) = \int P(\mathcal{D} | f) P(f)~\mathrm{d}f$ is given by
\begin{align}
    \log P(\mathcal{D}) =&-\frac{1}{2} \bm{y}^{\top}\left(K(\bm{x},\bm{x})+\sigma_\mathrm{n}^2 \bm{I}\right)^{-1} \bm{y} \nonumber\\
    &-\frac{1}{2} \log \left|K(\bm{x},\bm{x})+\sigma_\mathrm{n}^2 \bm{I}\right|-\frac{N}{2} \log 2 \pi.
    \label{eq:gpmarginalloglikelihood}
\end{align}

\subsection{Kernel structure learning}
\label{sec:structurelearning}
While GPs offer closed-form solutions for both the predictive distribution and the marginal likelihood, the challenge lies in determining the appropriate kernel function for a given task. In practise, one often resorts to selecting a kernel from a standard set of pre-defined kernels, and optimising their hyperparameters based on the available data, or sampling them using methods such as MCMC. We provide a list of common kernels and hyperparameter optimisation techniques in Appendix \ref{sec:kernels}. 

Still, there is often no clear recipe for the choice of kernel. While physical arguments are sometimes a helpful guide (e.g. \citet{Foreman-Mackey2017} argue the Matérn family of kernels is useful for modelling stellar variability), or expected properties of the function space (e.g. smoothness, stationarity, etc.) provide direction, kernel selection is usually performed manually, which is time-consuming and prone to biases.

\begin{figure}[t]
    \centering
    \includegraphics[width=0.5\columnwidth]{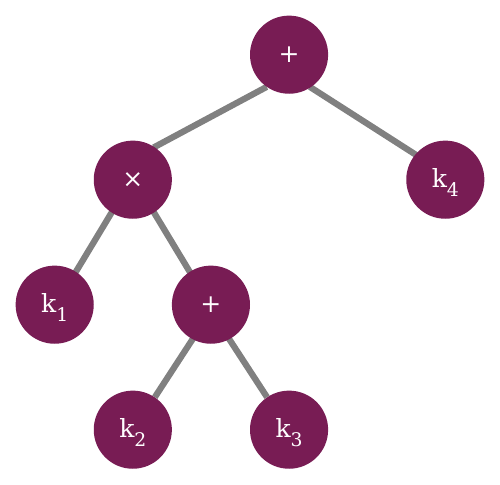}
    \caption{Example of a combinatorial kernel and its binary tree representation. The kernel, $k,$ is described by a non-linear combination of four elementary kernels, 
    $k = k_1 \cdot (k_2 + k_3) + k_4$. This combination can be visualised as a binary tree, where the leaf nodes correspond to the elementary kernels (atoms), and the internal nodes represent operations.}
    \label{fig:kernel_tree}
\end{figure}

In contrast, structure learning aims to learn the appropriate kernel structure, alongside its hyperparameters, directly from the data. Several approaches have been developed to address this challenge. A basic insight is that combinations of kernel functions, through addition and multiplication, also yield valid kernel functions. This allows for the systematic exploration of kernel structures by compositionally combining elementary kernels from a predefined library. The structure of the resulting kernel can be visualised as a full binary tree (Fig. \ref{fig:kernel_tree}), where the leaves represent elementary kernels (hereafter referred to as atoms), and the internal nodes represent operations (addition or multiplication). The task of structure learning then becomes identifying a suitable tree structure for the problem at hand.

One approach, introduced by \citet{Duvenaud2013} and further developed by \citet{Kim2018}, uses an iterative, greedy strategy. This process typically begins with a simple base kernel. A GP is then fitted to the data using this initial kernel, and the goodness-of-fit is evaluated. If the fit is deemed suboptimal, a new, more complex kernel structure is explored by either adding an elementary kernel to, or multiplying an elementary kernel with, the existing structure. In the context of the binary tree representation, this corresponds to expanding the tree by adding new leaves or modifying existing ones. The quality of fit for all kernel structures within a given layer of complexity is assessed using an objective function. The structures exhibiting the best fit are then propagated forwards to the next layer, iteratively increasing complexity, until no further improvement in the objective is observed. Common objectives used in this context include the Akaike information criterion $\text{AIC} = 2k - 2 \ln(\hat{L})$ \citep{Akaike1998},
and the Bayesian information criterion $\text{BIC} = k \ln N - 2 \ln(\hat{L})$ \citep{Schwarz1978a}, where $N$ is the number of parameters, $k$ is the number of hyperparameters in the model and $\hat{L}$ is the likelihood. A detailed implementation of this procedure can this process can be found in Algorithm 2 of \citet{Kim2018}.

While this iterative approach comes with the advantage of easy interpretability and mimics, in a sense, the manual kernel construction process a researcher would undertake, it has a number of drawbacks. Namely, the optimisation can get stuck in local optima during the search process, the optimisation can be unstable, and the procedure is prone to overfitting, especially when the data are sparse.

A more robust alternative is to treat kernel structure learning within a fully Bayesian framework. Rather than specifying a single kernel, we can construct a prior distribution over different kernel structures and their associated hyperparameters. The posterior distribution for the latent function $f$, considering the kernel structure prior, is then given by\begin{equation}
    P(f | \mathcal{D}) ~\propto~ \mathcal{L}(\mathcal{D} | f) \cdot P(f | k, \bm{\eta}) \cdot P(k, \bm{\eta}),
    \label{eq:gpposteriorwithkernelprior}
\end{equation}
where $P(k, \bm{\eta})$ represents the prior distribution over kernel structures $k$ and their hyperparameters $\bm{\eta}$. The binary tree representation of kernel structures naturally suggests a recipe for defining the kernel structure prior. For instance, each node type (addition, multiplication, atom) and each atom type can be associated with a prior probability.  The prior probability for a specific tree structure can be derived by multiplying the probabilities of its constituent nodes and atoms. From this perspective, larger, more complex tree structures, composed of more operations and atoms, inherently have a lower prior probability than simpler trees, which naturally helps mitigate overfitting. We can then, in principle, use Eq. \ref{eq:gpposteriorwithkernelprior} to sample latent functions and calculate posterior probabilities.

This Bayesian framework comes with some computational challenges.  Specifically, the kernel structure prior is defined over a discrete and combinatorial space, which makes standard MCMC algorithms, designed for continuous parameter spaces, inapplicable. Furthermore, the problem is trans-dimensional, as the dimensionality of the continuous hyperparameter vector $\bm{\eta}$ increases with the complexity of the kernel structure. We address these challenges by employing a custom SMC-based sampling algorithm, as detailed by \citet{Saad2023}, implemented in \gallifrey (see Sect. \ref{sec:gallifrey}).

\subsection{Sequential Monte Carlo}
\label{sec:smc}
The posterior in Eq. \ref{eq:gpposteriorwithkernelprior} can be sampled using any sampling algorithm that is adapted to the peculiarities of this specific problem formulation (i.e. a partially discrete, partially continuous trans-dimensional parameter space). In particular, one could directly use an adapted MCMC algorithm, as was done by \citet{Saad2019}. However, with the specific focus on time series modelling, SMC sampling offers some distinct advantages. SMC is closely related to MCMC, and can be understood as a population-based approach to Monte Carlo sampling.

At its heart, an SMC algorithm approximates a target probability distribution by iteratively refining an ensemble of weighted samples, referred to as particles. 
The goal of SMC is to sample from a sequence of distributions, $\{\pi_t\}_{t=1}^T$, where $\pi_t$ gradually approaches the desired target posterior distribution $P(f | \mathcal{D})$. We began by sampling an initial set of particles from a simpler distribution $\pi_0$, usually the prior. Subsequently, these particles were evolved through a series of steps to represent $\pi_1, \pi_2, \dots, \pi_T \approx P(f | \mathcal{D})$.

The evolution of particles in SMC involves three steps, which are repeated for each transition $\pi_{t-1} \rightarrow \pi_t$ (for $t=1, 2, \dots, T$):

\begin{enumerate}
    \item \textbf{Re-weighting}:  In each iteration, the particles are assigned weights that reflect how well they align with the next distribution in the sequence, relative to the previous one. Particles that match the new distribution more closely
    receive higher weights, effectively adjusting their importance in the ensemble. Given a set of $N$ particles $\{x_i^{(t-1)}\}_{i=1}^N$ approximating the distribution $\pi_{t-1}$ with normalised weights $\{w_i^{(t-1)}\}_{i=1}^N$, the updated un-normalised weights are calculated as
    \begin{equation}
        w_i^{*(t)} = w_i^{(t-1)} \frac{\tilde{\pi}_t(x_i^{(t-1)})}{\tilde{\pi}_{t-1}(x_i^{(t-1)})},
        \label{eq:reweighting}
    \end{equation}
    where $\tilde{\pi}_t$ and $\tilde{\pi}_{t-1}$ are the un-normalised target densities for distributions $\pi_t$ and $\pi_{t-1}$, respectively. From the un-normalised weights $w_i^{*(t)}$, the normalised weights $w_i^{(t)} = w_i^{*(t)}/\sum_i w_i^{*(t)}$ are calculated.
    
    \item \textbf{Resampling}: After re-weighting, some particles can have very low weights, barely contributing to the approximation, while others have disproportionately high weights, leading to sample degeneracy.  Resampling addresses this by eliminating particles with low weights and duplicating those with high weights. The simplest resampling technique is multinomial resampling, where a new set of $N$ particles are drawn from the ensemble $\{x_i^{(t-1)}\}_{i=1}^N$, where the probability of drawing the $i$-th particle is equal to the weight $w_i^{(t)}$ (with replacement). After the resampling, all particle weights are set to $w_i^{(t)} = 1/N$. Particles with higher weights in the previous step are more likely to be selected multiple times, effectively concentrating the samples in regions of higher probability mass.
    
    \item \textbf{Rejuvenation}:  To increase particle diversity after resampling (which introduces duplicates), a rejuvenation step is applied.  Typically, this involves applying a number of MCMC steps to move the particles to new locations. This step increases particle diversity and enhances exploration of the sample space, leading to an improved posterior estimate. 
\end{enumerate}

In the context of time series structure learning, the sequence of distributions $\{\pi_t\}_{t=1}^T$ can be constructed using data annealing \citep[e.g.][]{Karamanis2025}.  Starting with the prior distribution $\pi_0$ over model parameters (kernel structures and hyperparameters), each step $t$ introduces a new batch of observational data. The distributions $\pi_t$ then approximates the posterior distribution given the data observed up to step $t$. We discuss a number of common convergence tests for SMC in Appendix \ref{sec:smc_assessment}.

Sequential Monte Carlo has a number of advantages over MCMC for the GP time series structure learning context. The resampling step concentrates more computational effort in high probability regions, thereby increasing the quality of the approximation. Furthermore, GPs are known to scale very unfavourably, exhibiting a $O(N^3)$ scaling behaviour in the number of data points. By sequentially introducing additional data points, the inference can be sped up significantly. 

\section{The \gallifrey package}
\label{sec:gallifrey}

To make GP time series structure learning methods approachable for astronomical applications, especially in the context of exoplanet light curve modelling, we developed the \gallifrey Python package. This package implements the SMC algorithm with a Bayesian prior over kernel structures, as originally described by \citet{Saad2023} and inspired by the \texttt{AutoGP.jl} Julia package. The package is built using the JAX framework \citep{jax2018github} to ensure computational efficiency and scalability. JAX offers native parallelisation capabilities, which can be leverages for an inherently parallel algorithm like SMC. Furthermore, JAX's automatic differentiation simplifies the implementation of gradient-based sampling methods like Hamiltonian Monte Carlo (HMC), which is used for particle rejuvenation.

In this section we give a general overview of the inner workings of \gallifrey and some of its design choices. For a more in-depth description of the algorithmic details, see \citet{Saad2023}.

\subsection{Prior over kernel structures}
The kernel structure prior is defined over a space of possible kernel functions, constructed through a grammar of kernel composition. The package employs a library of atomic kernels, $\mathcal{K} = \{k_1, k_2,\dots, k_n\}$, and a set of operators, $\mathcal{O} = \{O_1, O_2,\dots, O_m\}$. These components are defined within the \texttt{GPConfig} class, allowing for user customisation and extension. Natively, \gallifrey supports addition and summations operators. A list of implemented atomic kernels can be found in \cref{tab:gpkernels}. 

Kernel structures are sampled recursively. At each node in the kernel tree, a choice is made between selecting an atomic kernel or an operator, based on predefined probabilities $p_\mathrm{kernel}$ and $p_\mathrm{operator}$ for each component. The probability of sampling a specific kernel structure $K$ is thus determined by the product of the probabilities of selecting each component in its construction:

\begin{equation}
P(K) = \prod_{\mathrm{node}~\in~K} P(\text{component at node})
.\end{equation}
This prescription naturally favours simple kernel structures over more complicated ones. The kernel construction process is constrained by a maximum tree depth, $D_\mathrm{max}$, to ensure computational tractability and prevent overfitting. The prior probabilities $p_\mathrm{kernel}$ and $p_\mathrm{operator}$ act as hyperparameters, allowing users to guide the search towards kernel structures of desired complexity.

\subsection{Priors for kernel hyperparameters}
Each atomic kernel $k \in \mathcal{K}$ is parameterised by a set of hyperparameters, $\bm{\eta}_k$. To complete the Bayesian model specification, we defined hierarchical priors over these hyperparameters, i.e. $P(k, \bm{\eta}) = P(\bm{\eta}|k) P(k) $. In \gallifrey, we employed weakly informative priors, typically log-normal distributions for positive hyperparameters (e.g., length scales and variances) and logit-normal distributions for bounded hyperparameters (e.g. powers and sigmoid parameters). The variance of the likelihood function $\sigma_n^2$ can be specified to a fixed value, if it describes for example observational noise. Alternatively, $\sigma_n^2$ can be sampled, in which case the prior is described by an inverse-gamma distribution. 

\subsubsection{Sequential Monte Carlo inference engine}

\gallifrey employs SMC to perform Bayesian inference over the space of kernel structures and their hyperparameters. The SMC algorithm iteratively refines a population of particles, each representing a candidate kernel structure and parameter set, as more data are sequentially incorporated.

\paragraph{Re-weighting:}  As new data points are added, particle weights are updated based on their marginal log-likelihood. The weight update for particle $i$ at SMC step $t$ is proportional to the ratio of marginal likelihoods:

\begin{equation}
w_i^{(t)} ~\propto~ w_i^{(t-1)} \frac{P(\mathbf{y}_t | \mathbf{x}_t, K_i, \boldsymbol{\eta}_i)}{P(\mathbf{y}_{t-1} | \mathbf{x}_{t-1}, K_i, \boldsymbol{\eta}_i)}
,\end{equation}

\noindent where $\mathbf{X}_t$ and $\mathbf{y}_t$ represent the data up to time $t$, and $K_i$ and $\boldsymbol{\eta}_i$ are the kernel structure and hyperparameters for particle $i$. The marginal log-likelihood is given by Eq. \ref{eq:gpmarginalloglikelihood}.

\paragraph{Resampling:}To avoid particle degeneracy, a resampling step is performed when the normalised effective sample size (ESS), 
\begin{equation}
\text{ESS} = \frac{1}{\sum_{i=1}^N \cdot \left(w_i^{(t)}\right)^2}
\label{eq:ESS}
,\end{equation}

\noindent falls below a predefined threshold (by default $1/2 \cdot N$, where $N$ is the number of particles). \gallifrey utilises stratified resampling, implemented through the \texttt{BlackJAX} package, by default, which reduced the variance of the resulting sample compared to simple multinomial resampling.

\paragraph{Rejuvenation:} After the resampling, a number of rejuvenation moves are performed. The rejuvenation involves a hybrid MCMC strategy:
\begin{enumerate}
    \item \textbf{Structure MCMC move:} A new kernel structure, $K_i'$, is proposed for each particle, $i,$ by modifying the binary tree structure describing the kernel. These moves are accepted or rejected based on a Metropolis-Hastings criterion, considering the kernel prior $P(K)$ and the marginal likelihood $P(\mathbf{y}_t | \mathbf{X}_t, K, \boldsymbol{\theta})$.
    \item \textbf{Parameter HMC move:} If the structure move is accepted, the new continuous hyperparameters $\boldsymbol{\eta}_i'$ of the new kernel structure $K_i'$, and optionally a new noise variance $\sigma_n^2$, are proposed by applying a number of HMC steps in the continuous parameter space. The HMC algorithm is implemented using the Python sampling package \texttt{BlackJAX} \citep{Cabezas2024}.
\end{enumerate}

Once all SMC rounds are complete, the algorithm returns a final sample of particles $\{x_i^{T}\}_{i=1}^N$ with weights  $\{w_i^{T}\}_{i=1}^N$, which constitute a posterior sample of the GP and can be used for inference. In particular, since each particle corresponds to a GP, with kernel $k_i$, hyperparameter $\bm{\eta}_i$, and a predictive distribution given by Eq. \ref{eq:gppredictivedistribution}, the predictive distribution of the whole ensemble is given by a Gaussian mixture model (GMM),\begin{equation}
    P\left(\bm{y}^* | \bm{x}^*, \bm{x}, \bm{y}\right) = \sum_{i=1}^N w_i \cdot P\left(\bm{y}^* | \bm{x}^*, \bm{x}, \bm{y}, k_i, \bm{\eta}_i\right),
    \label{eq:gppredictivemixturedistribution}
\end{equation}
where the weights $w_i$ are precisely the importance weights $\{w_i^{T}\}_{i=1}^N$ returned by the SMC sampler.

\section{Example applications}
\label{sec:examples}
To demonstrate the usefulness of GP structure learning methods, we present a number of example applications in this section.

\subsection{Stellar variability}
\label{sec:stellar_variability}
\begin{figure}
    \centering
    \includegraphics[width=\columnwidth]{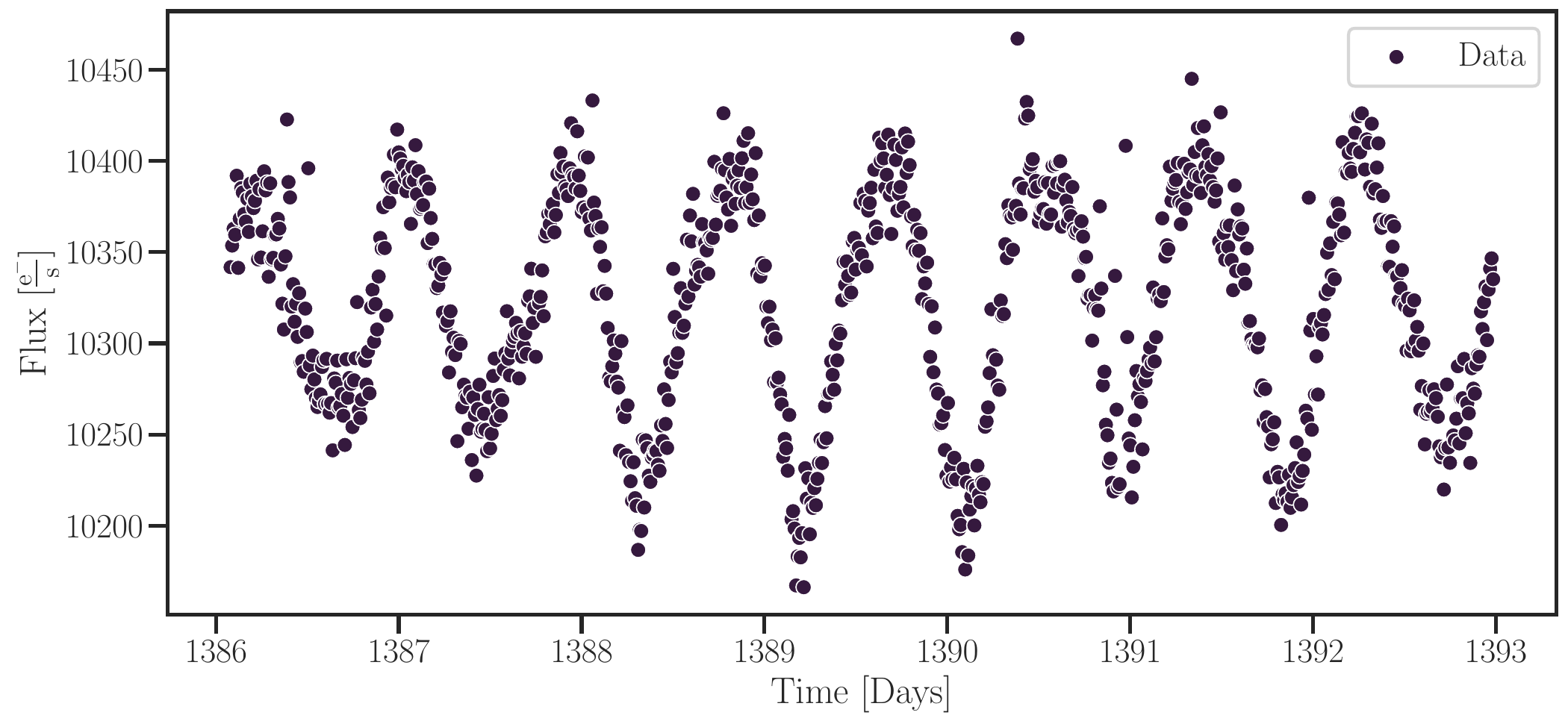}
    \caption{Example time series of stellar variability. The figure shows a seven-day window of the star TIC 10863087 observed with TESS, after removing outliers and thinning.}
    \label{fig:stellar_variability_data}
\end{figure}
We started the exploration by modelling the stellar variability of the star TIC 10863087, a variable M dwarf star observed by TESS. This target was previously used as a case study in the exoplanet modelling package \texttt{exoplanet} \citep{exoplanet:exoplanet}. The star has a mass of approximately $0.4~M_\odot$ and a rotation period of roughly $0.83$ days, rendering it a good test case for evaluating \gallifrey's capacity for interpolation and extrapolation of time series data. We retrieved an 8-day observation period from TESS using the \texttt{lightkurve} Python package \citep{2018ascl.soft12013L}. Following post-processing steps involving outlier removal and thinning, we obtained a time series comprising 814 data points, shown in Fig. \ref{fig:stellar_variability_data}.

\begin{figure}
    \centering
    \includegraphics[width=\columnwidth]{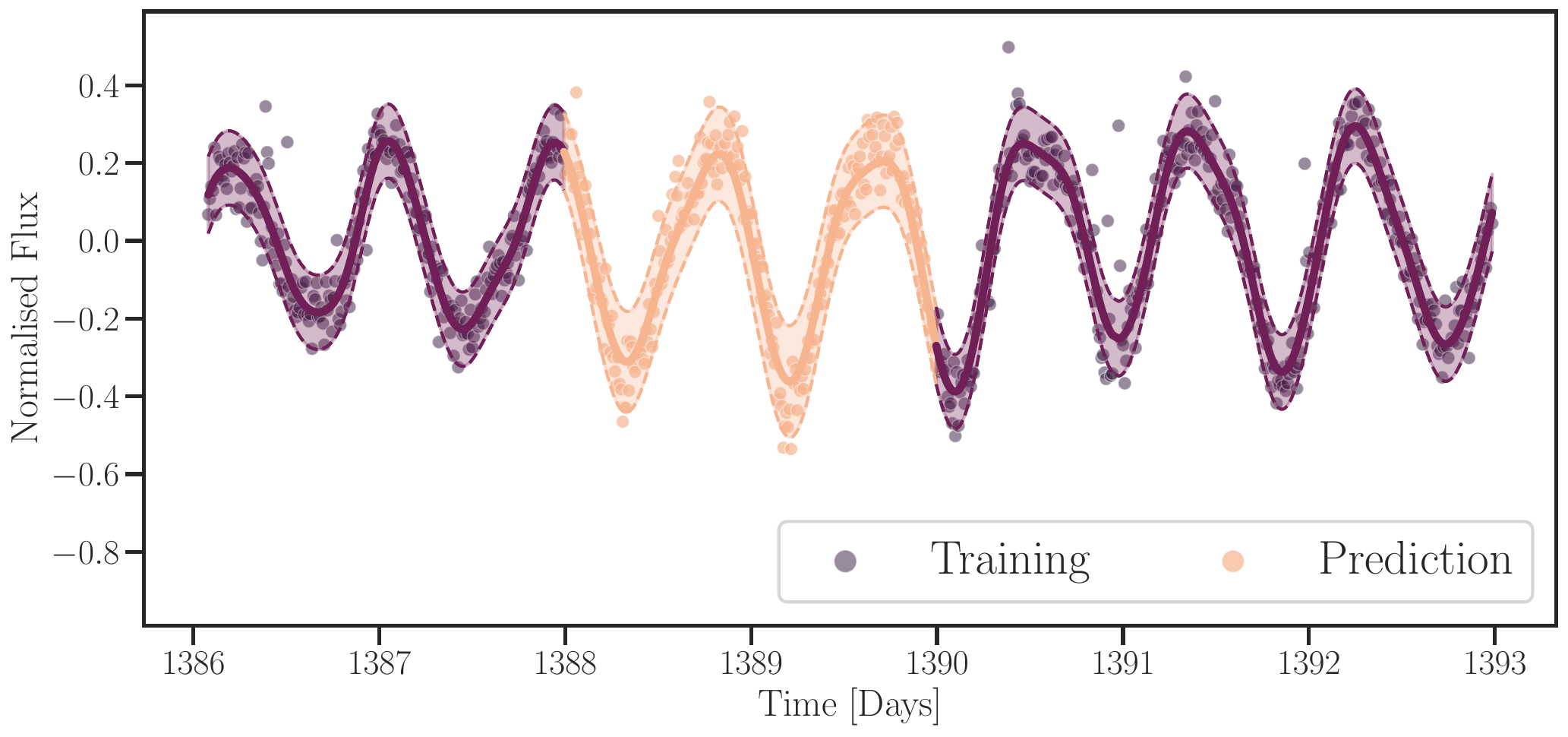}
    \caption{Demonstration of time series interpolation with \gallifrey. The figure shows the same data as Fig. \ref{fig:stellar_variability_data}. The purple data points have been used to train and condition the GP. The yellow region shows the prediction. The central line shows the mean prediction, and the confidence bands correspond to one standard deviation.}
    \label{fig:stellar_variability_interpolation_plot}
\end{figure}

\begin{figure*}
    \centering
    \includegraphics[width=\textwidth]{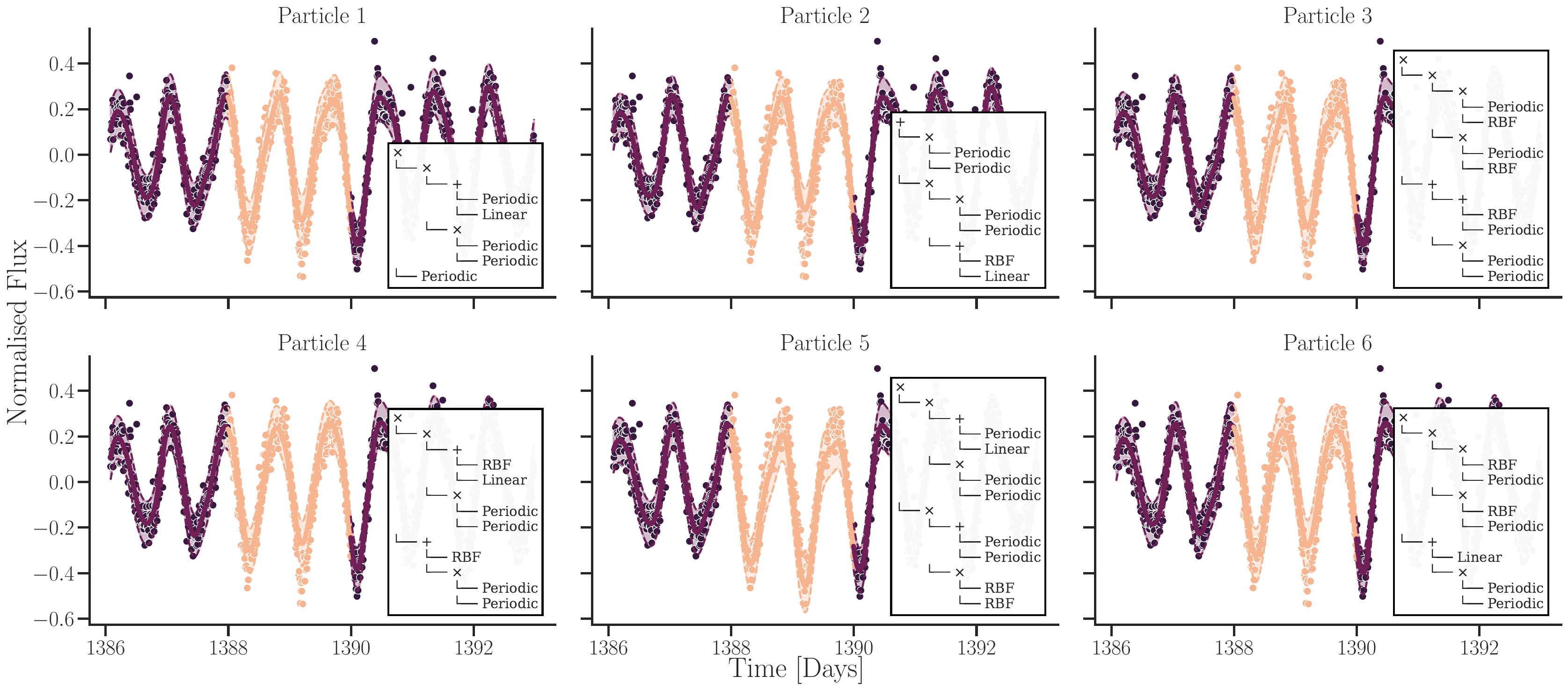}
    \caption{Sample of predictions for individual particles from the final SMC ensemble used for Fig. \ref{fig:stellar_variability_interpolation_plot}, including their respective kernel structures. Particles exhibit a variety of kernel structures and produce varying predictions, which leads to a more robust overall prediction when combined. Note that while the depth of every kernel displayed here corresponds to the maximum tree depth, $D_\text{max} = 3$, the learned kernels can end up shallower if simpler kernel structures are preferred. }
    \label{fig:stellar_variability_particle_interpolation}
\end{figure*}

To assess \gallifrey's interpolation capabilities, we masked a two-day window within the time series and train a GP on the remaining data. We employed standard \gallifrey settings, including a maximum tree depth of 3, and the linear, period, and radial basis function (RBF) kernels as atomic kernels. The training process utilises the SMC sampler with 64 particles and a linear annealing schedule, where approximately 5\% of data points are added in each round. We performed 75 structure MCMC moves and 10 parameter HMC moves per step. The resulting interpolation is shown in Fig. \ref{fig:stellar_variability_interpolation_plot}, showing a good match to the data. A key advantage of the SMC method is the availability of an entire ensemble of solutions, allowing us to examine individual predictions and the variety of learned kernel structures. A sample of particles including their learned kernel structures are shown in Fig. \ref{fig:stellar_variability_particle_interpolation}. We observe that \gallifrey identifies kernel structures that can become quite complex, while maintaining realistic uncertainties.

\begin{figure}
    \centering
    \includegraphics[width=\columnwidth]{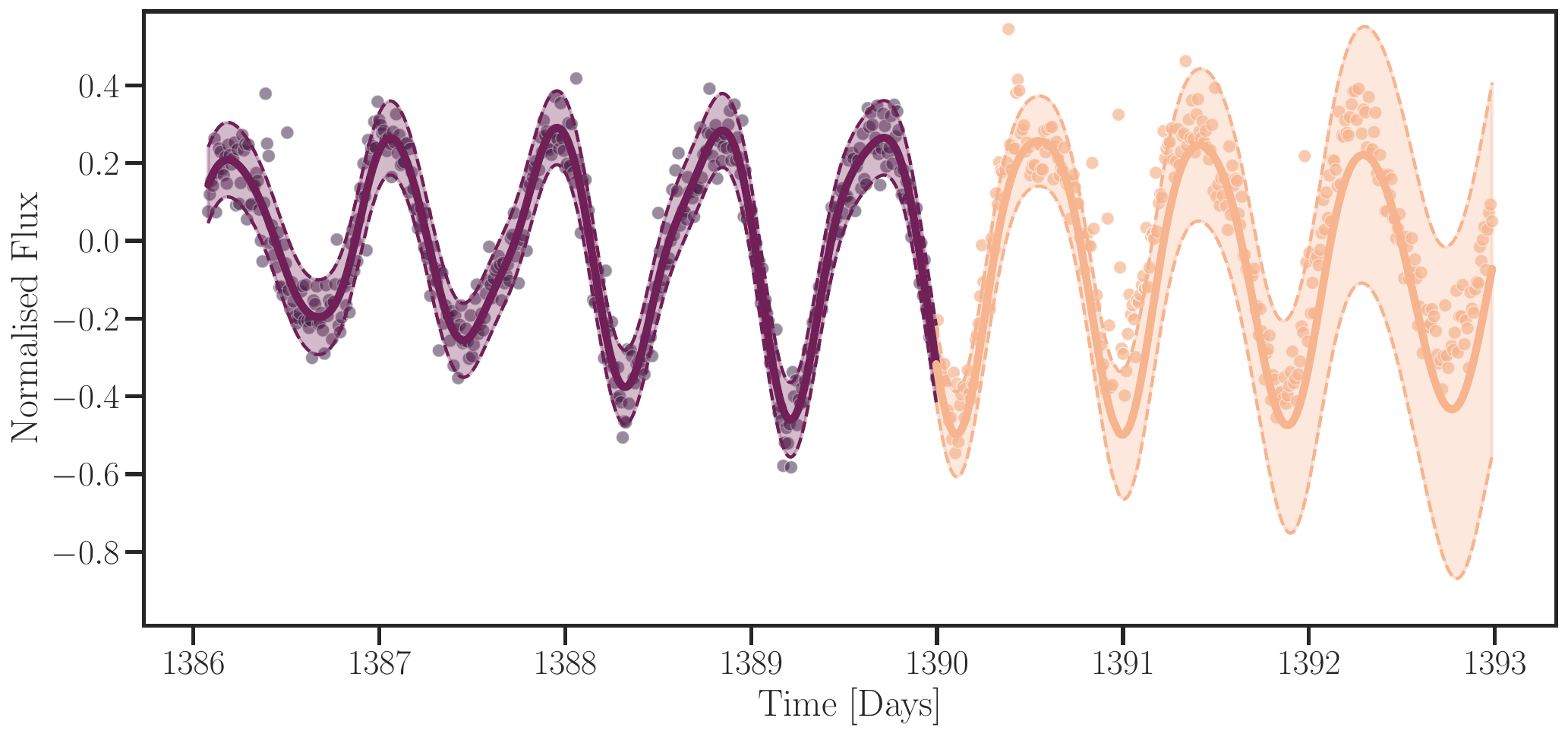}
    \caption{Same as Fig. \ref{fig:stellar_variability_interpolation_plot} but for forecasting rather than interpolation.}
    \label{fig:stellar_variability_extrapolation_plot}
\end{figure}

In addition to interpolation, we evaluated \gallifrey's forecasting capabilities. In this experiment, we included the first four days of data for training and attempted to forecast the subsequent three days, maintaining the same setup as before. The resulting forecast is presented in Fig. \ref{fig:stellar_variability_extrapolation_plot}. The forecast demonstrates good accuracy for up to one day. While deviations emerge afterwards, the uncertainty also increases, ensuring that the observed data remains well within the predicted uncertainty bounds. 

\begin{figure*}
    \centering
    \includegraphics[width=\textwidth]{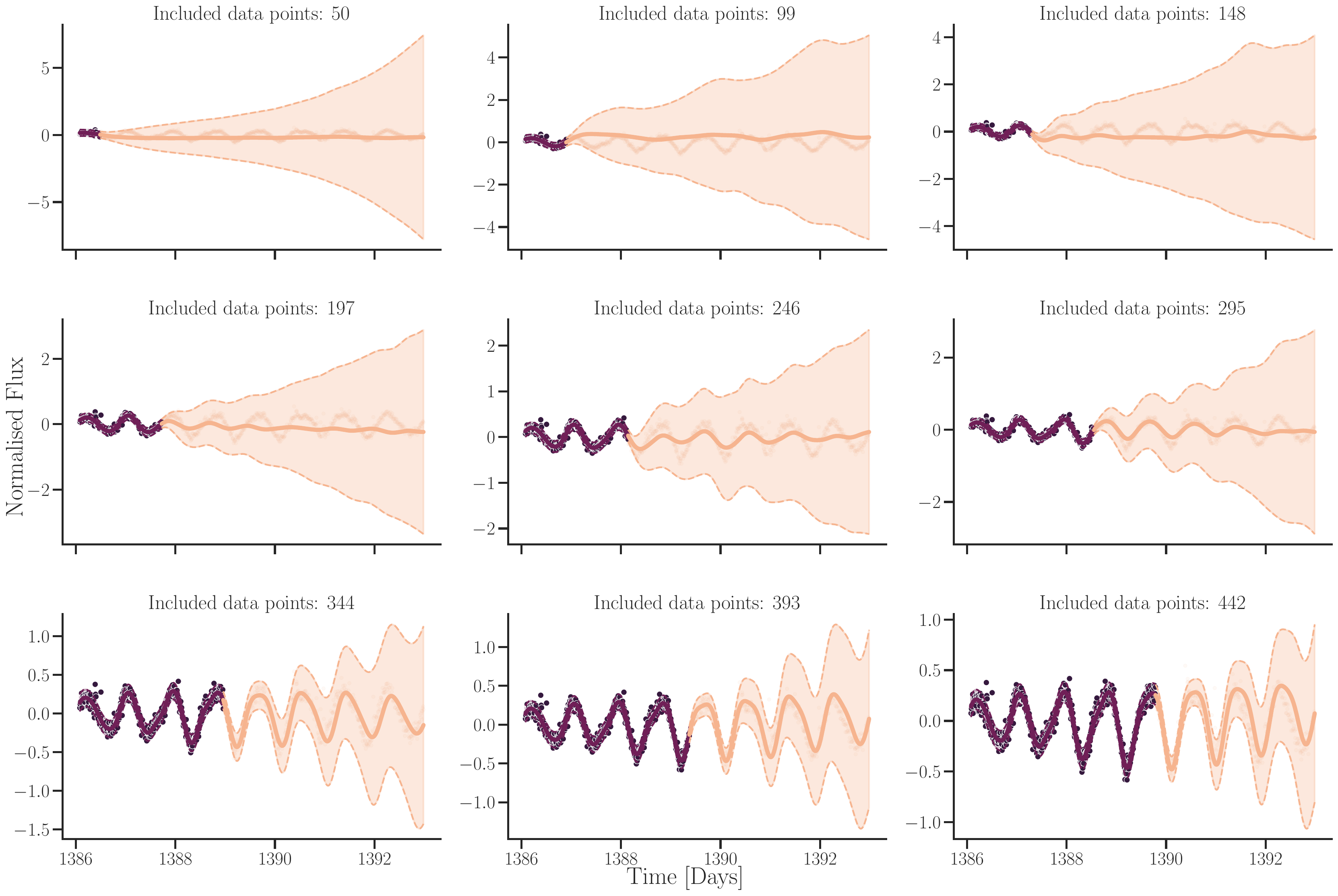}
    \caption{Snapshots of  SMC rounds during the structure learning process for producing the forecast shown in Fig. \ref{fig:stellar_variability_extrapolation_plot}. Purple data points have already been seen during the annealing schedule, while yellow regions are predictions. The algorithm learns more intricate patterns and reduces prediction uncertainty as more data points are added.}
    \label{fig:stellar_variability_extrapolation_history}
\end{figure*}

 Using the data annealing schedule inherent in SMC, we can visualise the evolution of the learned structure learning procedure. Figure \ref{fig:stellar_variability_extrapolation_history} visualises predictions after consecutive SMC rounds. Initially, with limited data, the prediction is highly uncertain, and the periodic pattern is not captured, leading to a predominantly flat forecast. As more data are incorporated, the prediction progressively improves, and periodic terms become increasingly dominant in the learned kernel structure.



\subsection{Transit fitting}
\label{subsec:transit_fit}
As seen in the previous section, the structure learning approach works well for time series interpolation, and this example demonstrates how to leverage these capabilities for transit parameter inference. For light curve fitting, GPs are frequently used as background models, and it is common to either jointly fit the light curve model $M(\bm{p})$, which is dependent on transit parameters $\bm{p}$, with the GP (effectively treating the transit model as a deterministic mean function in Eq. \ref{eq:gpdefinition}), or to utilise the GP for detrending. In this detrending approach, the GP is fit to the data, and the GP mean prediction is subtracted from the data.  Often, the transit region is masked out to prevent the GP from fitting the transit signal itself.

When employing learned kernels, simultaneous fitting of the light curve and the GP can introduce a significant risk of overfitting, especially if the GP's features operate on comparable timescales to the transit duration. On the other hand, simply subtracting the estimated GP mean leads to an underestimation of uncertainties and disregards correlations in the noise structure. To mitigate these issues, a hybrid approach can effectively separate the background and transit modelling steps in a two-stage process.

First, we divided the observational dataset $\mathcal{D} = \{\bm{t}, \bm{y}\}$ into two disjoint sets: $\mathcal{D}_\mathrm{trans} = \{\bm{t}_\mathrm{trans}, \bm{y}_\mathrm{trans}\}$, which contains the data points within the transit window, and $\mathcal{D}_\mathrm{bg} = \{\bm{t}_\mathrm{bg}, \bm{y}_\mathrm{bg}\}$, which contains the out-of-transit (background/training) data. We then applied the kernel structure learning procedure, from Sect. \ref{sec:methods}, exclusively to the background dataset $\mathcal{D}_\mathrm{bg}$. This allowed us to determine the optimal ensemble of kernel structures and their associated hyperparameters that best describe the noise and background model, without risking overfitting on the transit signal.

To estimate the transit parameters $\bm{p}$, we formulated the posterior distribution for $\bm{p}$ as
\begin{equation}
    P(\bm{p} | \mathcal{D}_\mathrm{trans}, \mathcal{D}_\mathrm{bg}) ~\propto~ P(\bm{y}_\mathrm{trans} | \bm{t}_\mathrm{trans}, \mathcal{D}_\mathrm{bg}, \bm{p}) \cdot P(\bm{p}),
    \label{eq:transit_posterior}
\end{equation}
where $P(\bm{p})$ is the prior distribution over the transit parameters. The likelihood term, $P(\bm{y}_\mathrm{trans} | \bm{t}_\mathrm{trans}, \mathcal{D}_\mathrm{bg}, \bm{p})$, is derived from the predictive distribution of the GP. Specifically, we assume that the observed in-transit data $\bm{y}_\mathrm{trans}$ can be modelled as the sum of a transit model $M(\bm{t}_\mathrm{trans}, \bm{p})$ and the background function, which is described by the GP. Therefore, the likelihood is given by
\begin{equation}
    P(\bm{y}_\mathrm{trans} | \bm{t}_\mathrm{trans}, \mathcal{D}_\mathrm{bg}, \bm{p}) = P(\bm{y}_\mathrm{trans} - M(\bm{t}_\mathrm{trans}, \bm{p}) | \bm{t}_\mathrm{trans}, \mathcal{D}_\mathrm{bg}).
\end{equation}
This likelihood is evaluated using the predictive mixture distribution obtained from the background GP model, given by Eq. \ref{eq:gppredictivemixturedistribution}. This approach leverages 
the full information on the time series structure learned by the SMC algorithm. By considering the learned covariance structure between data points, this approach also yields a more robust treatment of correlated noise compared to simple detrending methods.

\begin{figure}[ht]
     \centering
     \begin{subfigure}{\columnwidth}
        \centering
        \includegraphics[width=0.9\textwidth]{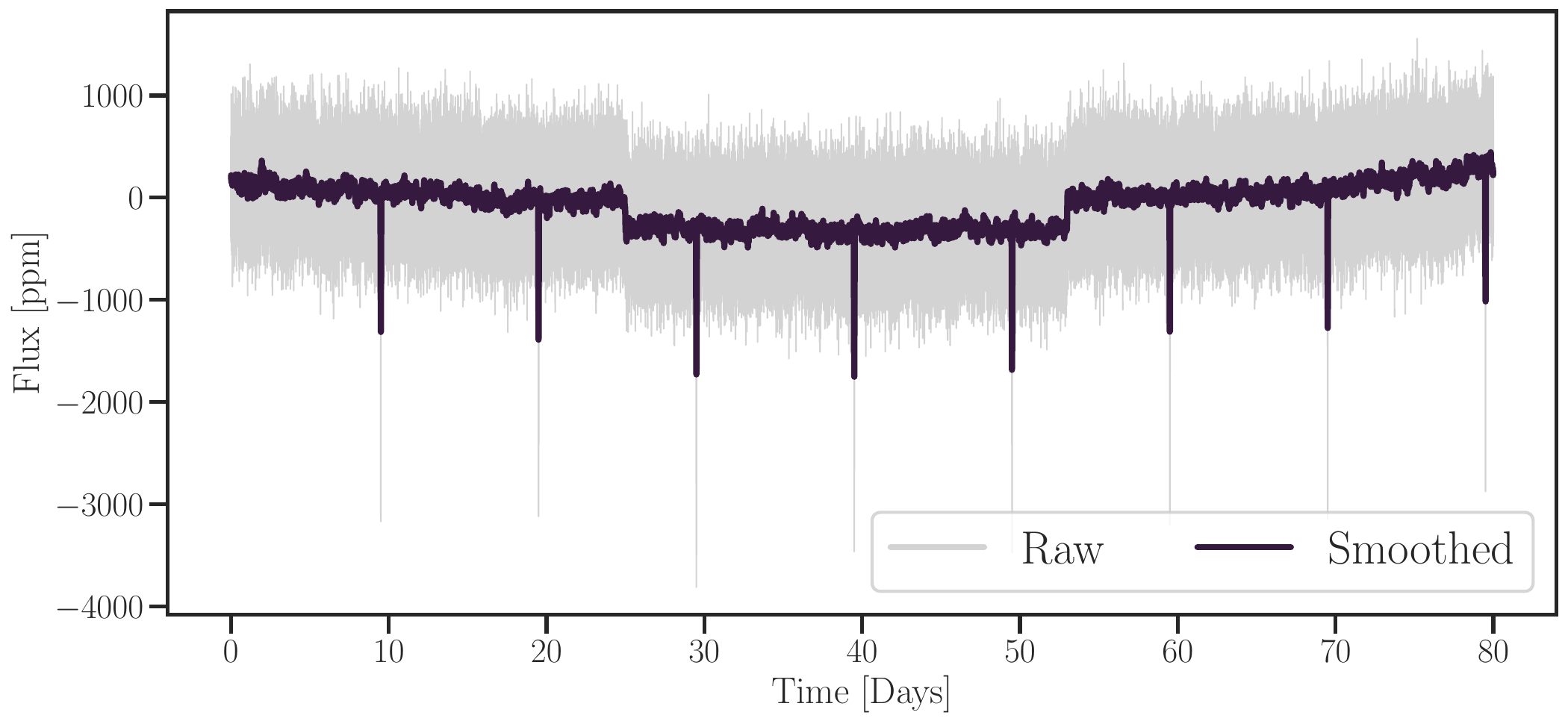}
        \caption{}
        \label{fig:plato_like_lightcurve}
     \end{subfigure}
     \begin{subfigure}{\columnwidth}
        \centering
        \includegraphics[width=0.9\textwidth]{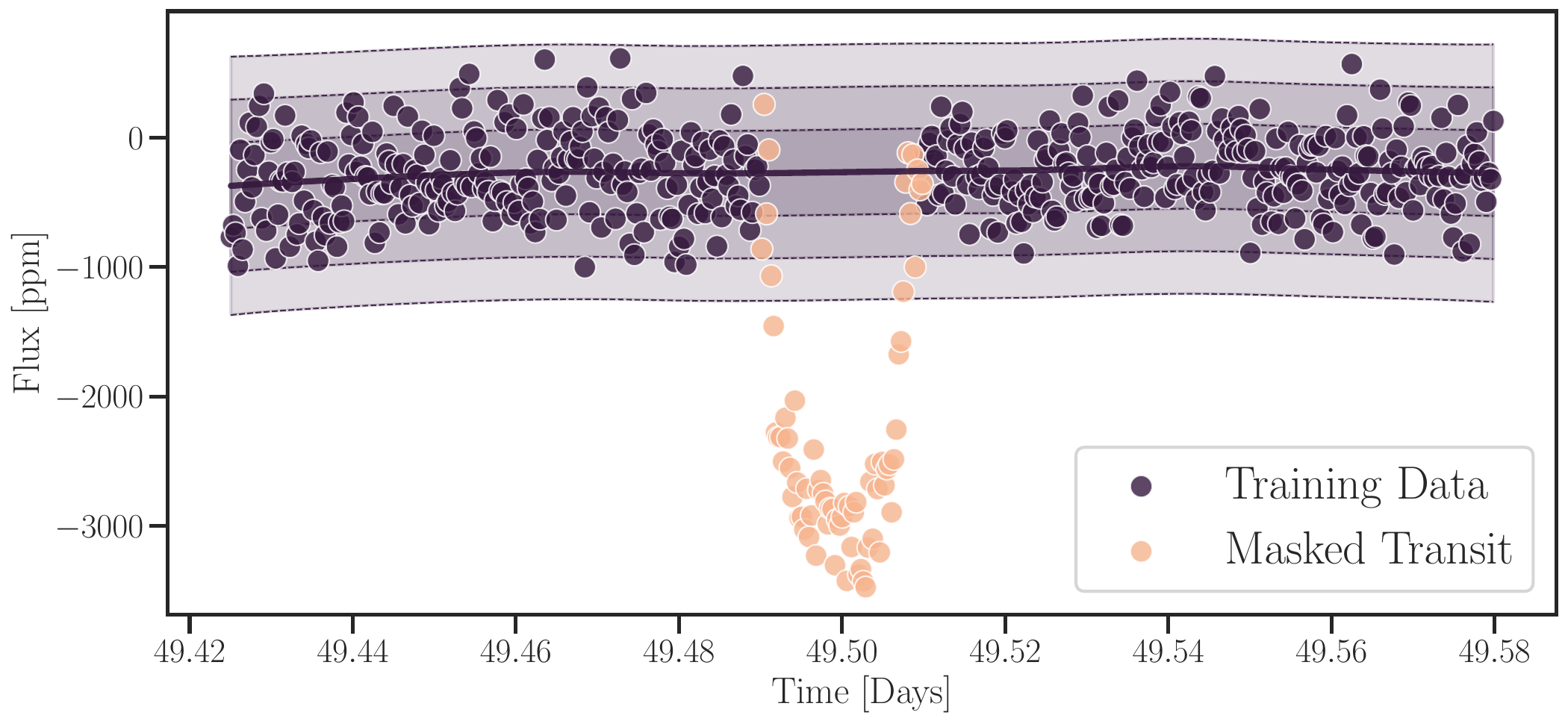}
        \caption{}
        \label{fig:plato_like_prediction}
     \end{subfigure}
     \begin{subfigure}{\columnwidth}
        \centering
        \includegraphics[width=0.9\textwidth]{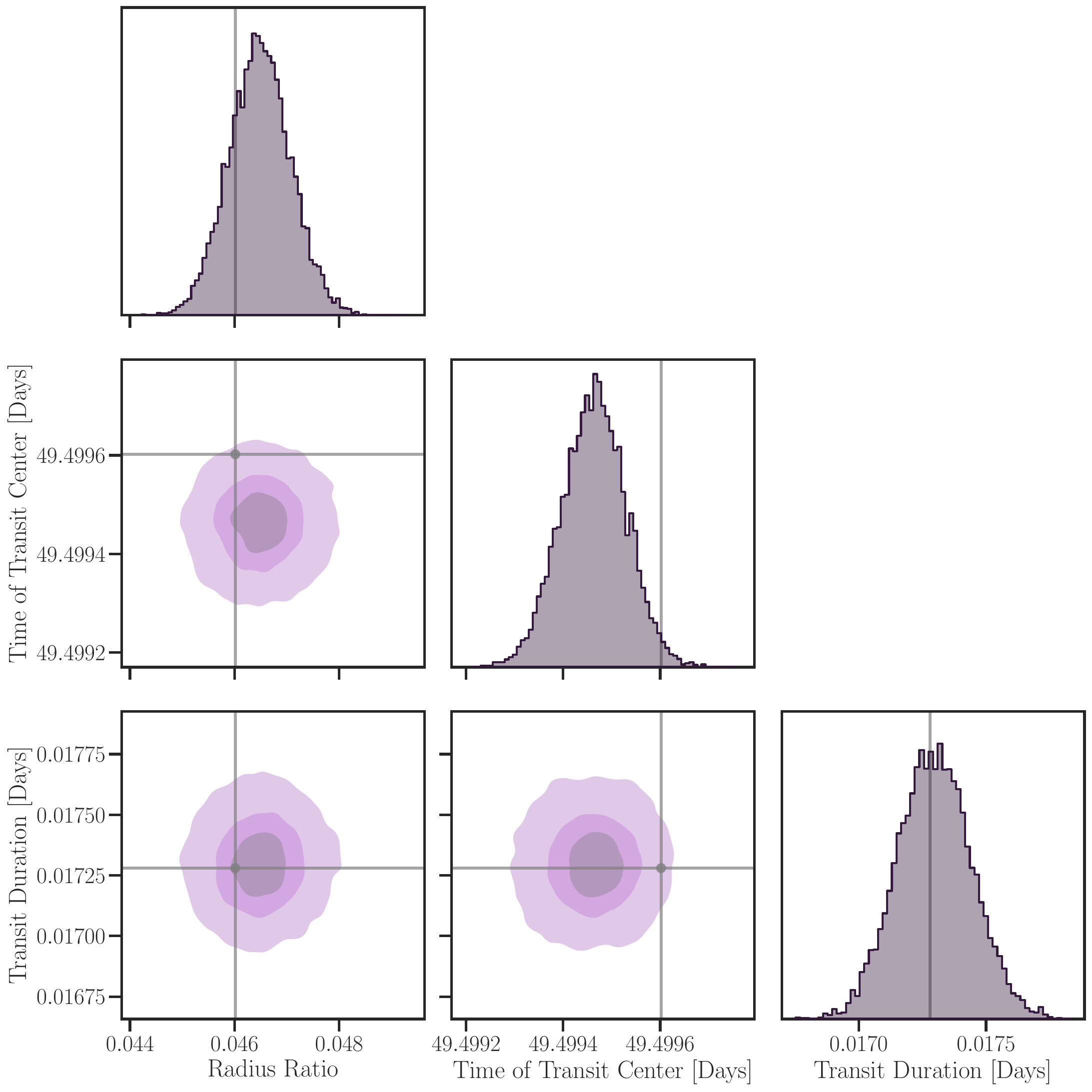}
        \caption{}
        \label{fig:plato_like_corner}
     \end{subfigure}
     \caption{Example of a transit parameter fit for a PLATO-like light curve. \textbf{(a)} Simulated 80-day light curve for 16 Cyg B using the \texttt{PLATO Solar-like Light-curve Simulator}. The grey curve shows the original light curve, and the purple shows a running mean average. \textbf{(b)} Zoom in on a single transit event. The purple data points are used to train and condition the GP, while the yellow data correspond to the masked transit. Also shown is the mean GP prediction, and the one, two, and three standard deviation confidence bands.
     \textbf{(c)} Posterior distribution for the transit parameter, $\bm{p},$ using the GP predictive distribution as likelihood. The grey lines represent the true values.}
     \label{fig:plato_like_lc_plots}
\end{figure}

\subsubsection{Fitting a transit for a PLATO-like light curve}
\label{sec:platofit}
To demonstrate this approach, we employed a synthetic light curve for the star 16 Cyg B (KIC 12069449), a main-sequence star with $M = 1.04~M_\odot$, generated using the \texttt{Plato Solar-Like Lightcurve Simulator} \citep{Samadi2019a}. We simulated an 80-day observation period and introduced a transit signal corresponding to a planet with a radius of $0.5~R_\mathrm{Jup}$ (Fig. \ref{fig:plato_like_lightcurve}). The parameters of the planet and star system are summarised in \cref{tab:plato_like_transit_parameters}.

For transit parameter estimation, we selected a time window encompassing a single transit event (Fig. \ref{fig:plato_like_prediction}), masked out the transit region, and trained \gallifrey to learn the background model using the out-of-transit data using an ensemble of 64 particles. Afterwards, we sampled the transit parameter posteriors using the predictive mixture distribution of the learned background model as the likelihood, as described above. We employed the JAX-based light curve modelling package \texttt{jaxoplanet} \citep{jaxoplanet} and the sampling package \texttt{BlackJAX} \citep{Cabezas2024}. For simplicity, we fixed the planet period, impact parameter, and limb darkening coefficients, and assumed flat priors over the remaining parameters.

The resulting posterior distribution for the transit parameters is shown in Fig. \ref{fig:plato_like_corner}. The planet-to-star radius radio and transit duration is recovered within one standard deviation, while the time of mid-transit is recovered within three standard deviations.

\begin{table}
    \centering
    \caption{Parameters of the simulated planetary system around 16 Cyg B and retrieved transit parameters.}
    \label{tab:plato_like_transit_parameters}
    \begin{tabularx}{\columnwidth}{LRR}
        Parameter & True value & Retrieved value \\
        \hline \hline
        \addlinespace
        Period [days] & 10 & --- \\
        \addlinespace
        Impact parameter & 0 & --- \\
        \addlinespace
        Limb darkening coefficients & [0.25, 0.75] & --- \\
        \addlinespace
        Radius ratio $R_p/R_\star$ & 0.046013 & $0.0465 \pm 0.0006$ \\
        \addlinespace
        Transit duration [days] & 0.01728 & $0.01730 \pm 0.00015$ \\
        \addlinespace
        Time of mid-transit [days] & 49.499602 & $49.49946 \pm 0.00007$ \\
        \addlinespace
        \hline
    \end{tabularx}
    \tablefoot{The retrieved values correspond to the mean and standard deviation of the MCMC chain. Parameters without retrieved values were fixed to their true value.}
\end{table}

\subsubsection{Comparison with a simple RBF kernel}
\label{sec:rbf_comparison}
The previous example demonstrated the effectiveness of the method on realistic transit data, making it suitable for current and upcoming transit missions like PLATO. However, the stellar background in that case was relatively simple, and a less complex kernel would likely have sufficed.  In this example, we aim to illustrate the difference between a structure learning and using a simple pre-defined kernel, when dealing with stellar variability and correlated noise.

We created a synthetic stellar background data with periodic, sinusoidal oscillations, and a non-linear trend. We also added correlated (red) noise, rather than white noise. This type of noise can be encountered, for example in ground-based observations, where the dynamical state of the atmosphere contributes to the observational noise.

Specifically, we employed an autoregressive model of order 1, AR(1), as the background noise model.  An AR(1) process is defined by

\begin{equation}
    \epsilon_t = \phi \cdot \epsilon_{t-1} + n_t,
\end{equation}

\noindent where $\epsilon_t$ is the noise at time $t$, $\phi$ is the autoregressive coefficient (with $0 \leq \phi < 1$), and $n_t$ is white noise with variance $\sigma_w^2$. This model introduces a dependence of the noise at time $t$ on the noise at the previous time step, $t-1$. The AR(1) process is the discrete-time analogue of the Ornstein-Uhlenbeck process, which has a covariance function given by $k(x, x') = \sigma^2 \exp(-|x - x'|/l)$, where $l$ is a length scale parameter related to $\phi$.  Previous work (e.g. \citealt{Pont2006}) has highlighted that neglecting this type of correlated noise can lead to biased estimates of transit parameters.

\begin{figure}[ht]
     \centering
     \begin{subfigure}{\columnwidth}
        \centering
        \includegraphics[width=0.9\textwidth]{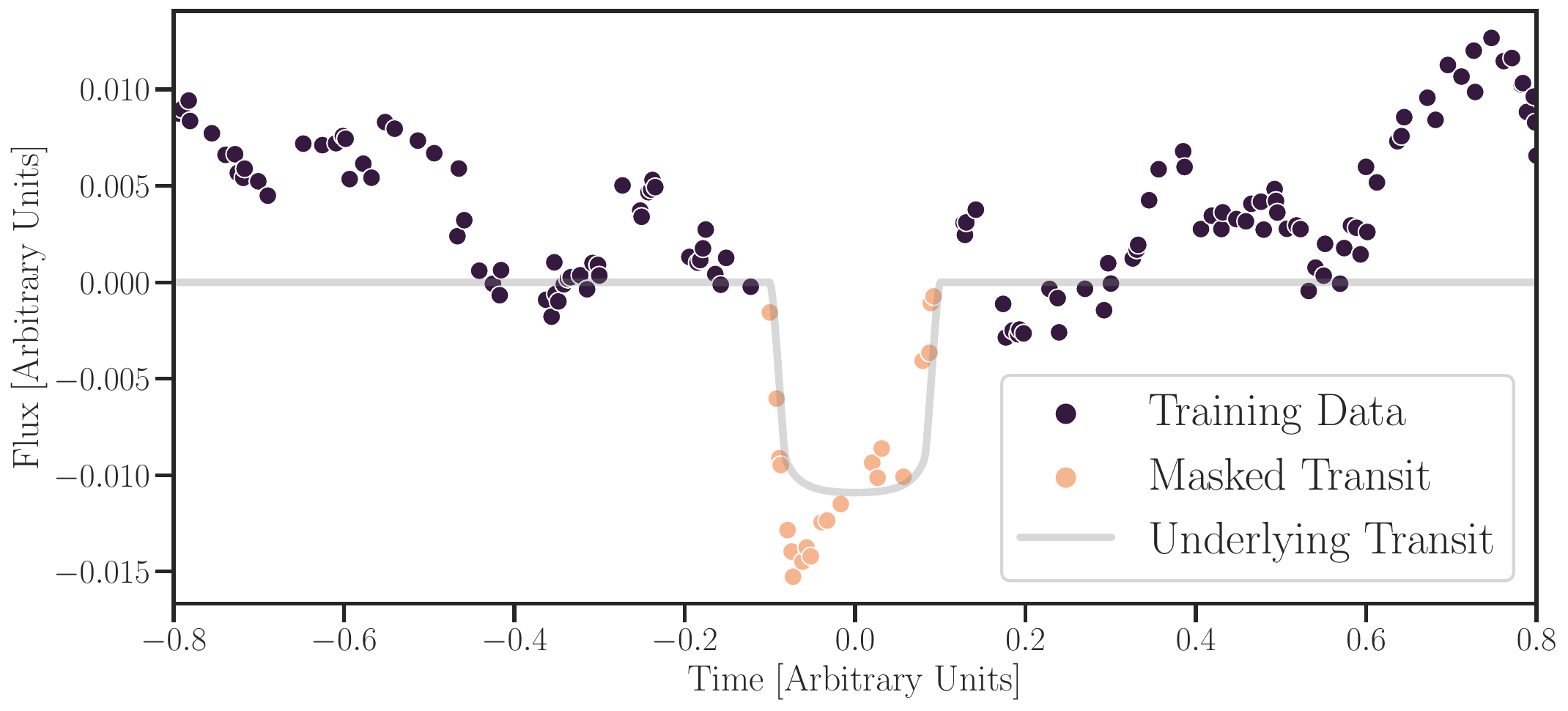}
        \caption{}
        \label{fig:kernel_comparison_data}
     \end{subfigure}
     \begin{subfigure}{\columnwidth}
        \centering
        \includegraphics[width=0.9\textwidth]{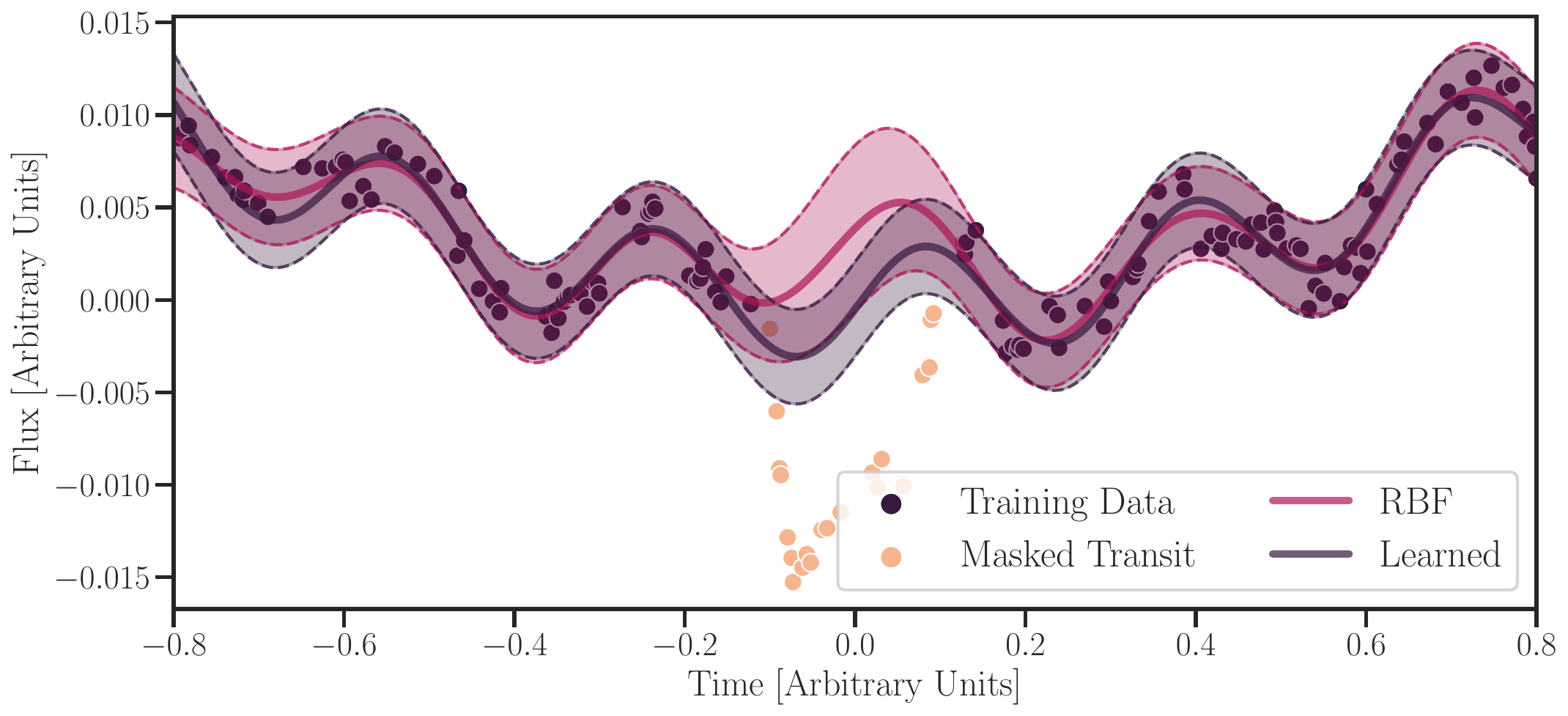}
        \caption{}
        \label{fig:kernel_comparison_prediction}
     \end{subfigure}
     \begin{subfigure}{\columnwidth}
        \centering
        \includegraphics[width=0.835\textwidth]{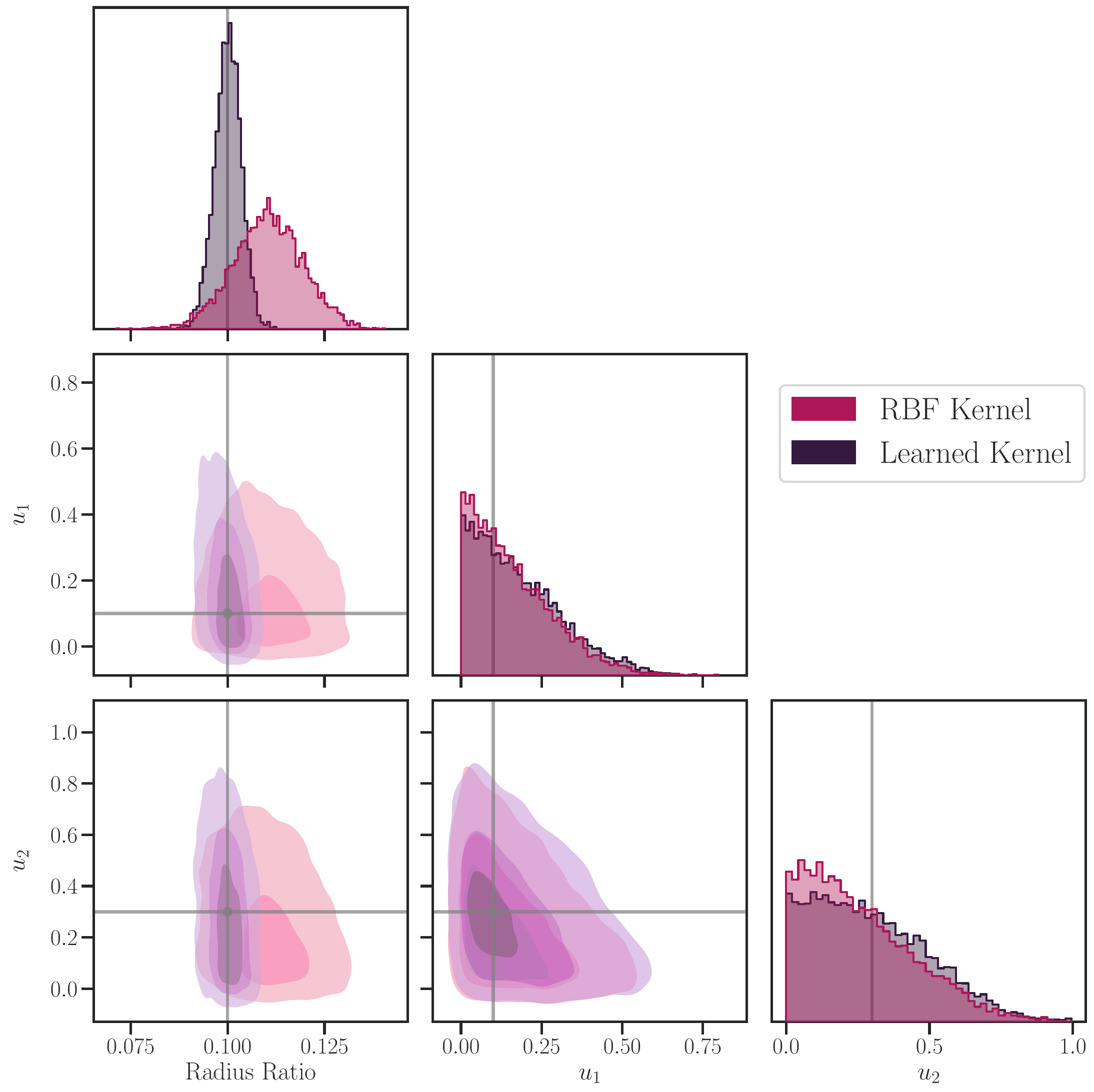}
        \caption{}
        \label{fig:kernel_comparison_corner}
     \end{subfigure}
     \caption{Transit parameter fit for the synthetic light curve with noticeable stellar variability and correlated noise in Sect. \ref{sec:rbf_comparison}. \textbf{(a)} Synthetic data with correlated noise generated using an AR(1) process, with an added transit signal. The grey line corresponds to the underlying transit light curve. \textbf{(b)}  GP prediction for the learned kernel structure (purple) and the RBF kernel (magenta). The central line corresponds to the mean prediction with a confidence band of one standard deviation. \textbf{(c)} Posterior distribution for the transit parameter, $p,$ using the learned kernel and RBF kernel to construct the predictive distribution. }
     \label{fig:kernel_comparison_plots}
\end{figure}

We added a transit signal to this synthetic data using \texttt{jaxoplanet}, the transit parameters are given in \cref{tab:comparison_transit_parameters}. The resulting light curve is shown in Fig. \ref{fig:kernel_comparison_data}.

We then applied the same two-stage approach as in Sect. \ref{subsec:transit_fit}. We masked the transit region and trained two separate GP models on the background data, $D_\text{bg}$. On the one hand, we used \gallifrey's structure learning with a maximum tree depth $D_\text{max} = 3$, and a kernel library consisting of the periodic, RBF, and linear kernels as atomic kernels. For comparison, we also trained a model with a maximum tree depth of $D_\text{max} = 0$ and only the RBF kernel as the base kernel. This mirrors typical applications of GPs, where a single, pre-defined kernel is chosen. The RBF kernel is a common default choice in such scenarios.

Figure \ref{fig:kernel_comparison_prediction} shows the learned GPs for both methods. The models perform similar outside the transit region, where training data are available. However, the predictions within the masked transit region differ significantly. The simple RBF kernel predicts a larger background value compared to the structure-learned model. This difference will directly impact the subsequent transit parameter estimation.

We sampled the transit parameter posteriors using the same approach as in the previous section. We fixed the planetary period, time of mid-transit, transit duration,and impact parameter to their true value, and fitted the planet-to-star radius ratio $R_p/R_*$ and the two limb darkening coefficients, $u_1$ and $u_2$. We again applied flat priors, where the limb darkening coefficients were constrained to lie between 0 and 1.

Figure \ref{fig:kernel_comparison_corner} shows the posterior distributions for both models. In both cases, the limb-darkening parameters cannot be meaningfully constrained and have similar distributions. However, the kernel learning method is able to constrain the radius ratio more precisely and more accurately. The mean value for the learned kernel is $0.100 \pm 0.004$, which matches the true value to three significant figures. In contrast, the RBF kernel method estimate is about one standard deviation away from the true value, while the standard deviation itself is more than twice as large for this kernel than for the learned kernel.

\begin{table}
    \centering
    \caption{Transit parameters of the simulated system in Sect. \ref{sec:rbf_comparison}.}
    \label{tab:comparison_transit_parameters}
    \begin{tabularx}{\columnwidth}{LRRR}
        Parameter & True value & RBF Kernel & Learned Kernel \\
        \hline \hline
        \addlinespace
        Period & 10 & --- & --- \\
        \addlinespace
        Time of mid-transit & 0.0 & --- & ---\\
        \addlinespace
        Transit duration & 0.2 & --- & --- \\
        \addlinespace
        Impact parameter & 0 & --- & ---\\
        \addlinespace
        Radius ratio & 0.1 & $0.110 \pm 0.009$ & $0.100 \pm 0.004$ \\
        \addlinespace
        $u_1$ & 0.1 & $<0.398$ & $<0.441$  \\
        \addlinespace
        $u_2$ & 0.3 & $<0.613$ & $<0.660$ \\
        \addlinespace
        \hline
    \end{tabularx}
    \tablefoot{In the RBF Kernel and Learned Kernel columns, the value for the radius ratio corresponds to the mean and standard deviation of the posterior MCMC chain. The upper limits for $u_1$ and $u_2$ correspond to the 95th percentile. Parameters without retrieved values were fixed to their true value. The period, time of mid-transit and transit duration are given in arbitrary units.}
\end{table}

\subsection{Transmission spectroscopy of HATS-46 b}
\label{subsec:transmission_spectroscopy}

As a final example, we used \gallifrey to obtain the transmission spectrum of HATS-46 b, a hot Jupiter with a Jupiter-like radius ($R_p = 0.95~R_{\text{Jupiter}}$) and a mass of $M_p = 0.16~M_{\text{Jupiter}}$. This planet has been observed using the EFOSC2 instrument on the ground-based New Technology Telescope, and the data have been analysed by \cite{Ahrer2023}.

In the original work, background modelling was performed using a variety of methods, including GPs with different kernels. Their analysis encompassed the white light curve and 25 spectroscopic light curves, and was performed in a two-step process. First, they fit the transit parameters of the white light curve and fixed the shared parameters $\bm{p}_\mathrm{shared}$ (stellar radius, inclination, and time of mid-transit) to their best-fit values. Subsequently, they sample the individual parameters $\bm{p}_\mathrm{individual}$ (radius ratio, and limb-darkening coefficients, $u_1$ and $u_2$) for each spectroscopic light curve independently.  This done to maintain consistency across the light curves, as different values of the stellar radius, for example, would lead to shifts in the derived transit depths relative to each other.

We used an alternative approach, building on the analysis techniques presented in the previous sections. We first fitted GP background models to all 26 light curves (the white light curve and 25 spectroscopic light curves), masking the transit regions during the fitting process. This provides, for each light curve, a GP model trained on the out-of-transit data.

From these trained GP models, we obtain 26 predictive distributions for the transit regions, one for each light curve. and then construct a transit model for each light curve, using the corresponding predictive distribution as the background model.  The likelihood for all parameters (shared and individual) is obtained by multiplying the individual likelihoods.  For a single light curve $i$, the likelihood is based on the GMM predictive distribution. The full likelihood is the product of all individual likelihoods, 

\begin{equation}
 \mathcal{L}(\bm{p}_{\text{shared}}, \{\bm{p}_{\text{ind},i}\}) =  \prod_{i=1}^{26} \mathcal{L}_i(\bm{p}_{\text{shared}}, \bm{p}_{\text{ind},i}).
\end{equation}

Our final model has 4 shared parameters: stellar radius ($R_\star$), inclination ($i$), stellar mass ($M_\star$), and time of mid-transit ($T_0$).  It also has $26 \times 3 = 78$ individual parameters: the radius ratio ($(R_p/R_\star)_i$), and the quadratic limb-darkening coefficients ($u_{1,i}$ and $u_{2,i}$) for each of the 26 light curves.  This gives a total of 82 parameters. We fixed the planet's orbital period to the value reported by \citet{Ahrer2023}, $T = 4.7423749$ days.

A key advantage of our approach is that we disentangle the GP fitting and the transit parameter fitting.  Traditionally, fitting both the GP hyperparameters and the transit parameters simultaneously would be computationally infeasible.  However, because we have already obtained a sample of GPs (and constructed their predictive distributions), we no longer need to sample the GP hyperparameters during the transit parameter estimation. This drastically reduces the computational cost.

Finally, we used efficient HMC sampling, specifically the No-U-Turn Sampler (NUTS; \citealt{Hoffman2011}) implemented in \texttt{BlackJAX}, to sample the 82 transit parameters simultaneously. This allows us to fully propagate the uncertainties in the shared parameters into the final transmission spectrum. 

\begin{figure}
    \centering
    \includegraphics[width=\columnwidth]{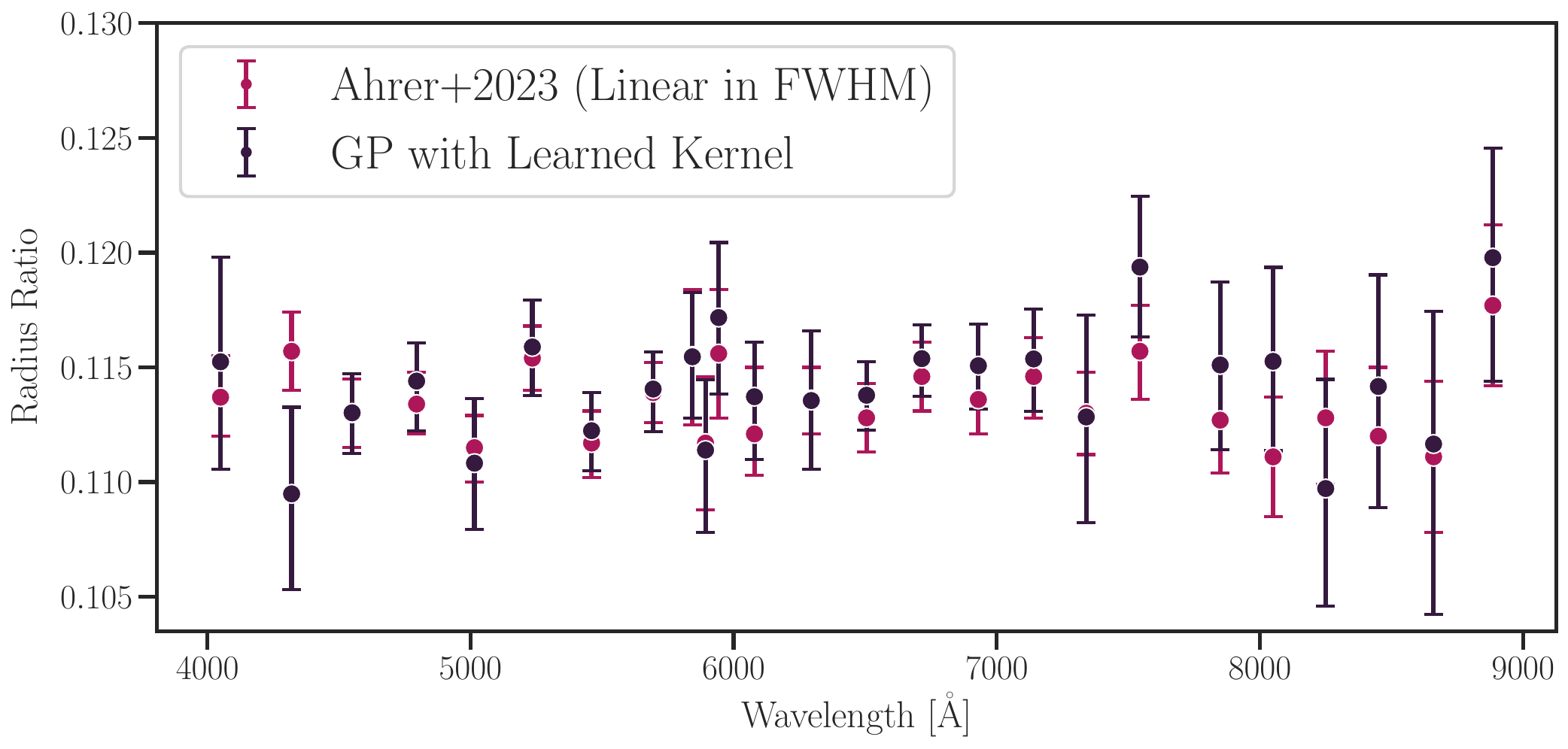}
    \caption{Transmission spectrum of HATS-46 b using \gallifrey's learned GP model (purple) and the reference spectrum obtained by \citet[magenta]{Ahrer2023}, using their fiducial detrending method.}
    \label{fig:HATS46b_transmission_spectrum}
\end{figure}

\begin{figure*}
    \centering
    \includegraphics[width=\textwidth]{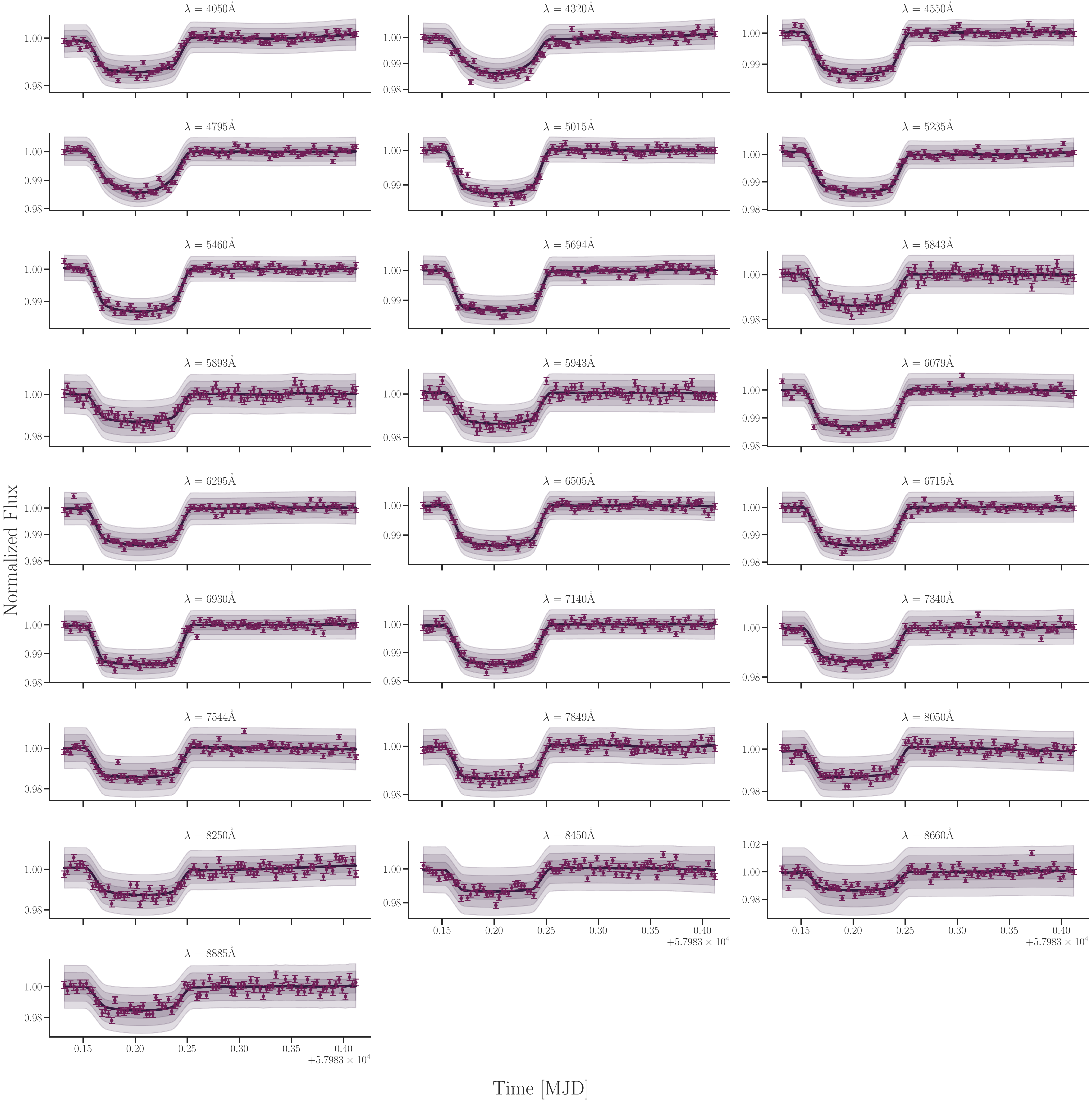}
    \caption{Model light curves for the 25 spectroscopic HATS-46 b light curves. Shown are the mean GP predictions with one, two, and three standard deviation confidence bands. The transit models are included using the median transit parameter from the MCMC sampling of the joint posterior.}
    \label{fig:HATS46b_light_curve_models}
\end{figure*}

The resulting transmission spectrum, derived from this joint analysis, is shown in Fig. \ref{fig:HATS46b_transmission_spectrum}, along with the spectrum from \citet{Ahrer2023} for comparison. Overall, the two spectra are consistent with one another, while our uncertainties are, on average, larger. This is expected because we included the uncertainties in the shared parameters (rather than fixing them), we incorporated different GP kernel structures, and we also sampled the second limb-darkening coefficient ($u_2$), which was fixed in the original analysis.  Despite these differences, we obtain a consistent and robust transmission spectrum because we accounted for all sources of uncertainty in a principled Bayesian manner. Figure \ref{fig:HATS46b_light_curve_models} shows the individual transit light curves and their corresponding fits, demonstrating the agreement between our model and the observed data. The model captures both the transit signal and the stellar variability, as represented by the GP predictive distributions.

\section{Discussion}
\label{sec:discussion}

We have demonstrated the potential of GP structure learning as a powerful tool for time series modelling tasks, particularly within the context of stellar activity and exoplanet transit modelling. We show that structure learning can lead to more robust, consistent, and in some cases more accurate transit parameter retrievals. We have implemented this structure learning approach in a self-consistent Bayesian manner, by constructing a prior over kernel structure that allows for the exploration of a space of plausible GP models. To ensure robust and efficient posterior sampling of this space, we used SMC methods, which benefit from native parallelisability and can create ensemble forecasts over different samples from the GP space. SMC also accelerates the sampling process, since it allows for the sequential addition of data points using a data annealing scheme, which partially offsets the inherently inefficient $O(N^3)$ scaling of GP regression. The combination of the structure learning method and SMC model ensembles enables the creation of robust predictive distributions, which allow us to decouple the background modelling task from the transit parameter inference. This results in a faster and more uncertainty-aware parameter retrieval while avoiding the risk of overfitting.

To encourage the wider adoption of these techniques, we have developed \gallifrey, a user-friendly Python package that encapsulates the described algorithms.  \gallifrey implements the structure learning methodology, originally described by \citet{Saad2023}, and is built on the JAX library \citep{jax2018github} and its surrounding ecosystem. This ensures computational efficiency and automatic differentiability, which is useful for gradient-based sampling methods in high-dimensional parameter spaces.  

We have shown the applicability of this approach to both space-based transit light curves, such as those from TESS or the upcoming PLATO mission, and ground-based spectroscopic observations. To demonstrate the advantages of disentangling the GP background and transit modelling, we give an example of obtaining the transmission spectrum of the hot Jupiter HATS-46 b from the spectroscopic light curves. By independently modelling the background in each light curve using flexible GPs trained on out-of-transit data, we generated predictive distributions for the in-transit background, which can be used as likelihood functions for the transit parameter retrieval. This allows for joint sampling of the transit parameters for all light curves, leading to more accurate and robust uncertainty estimates for the transmission spectrum.

Despite our focus on efficiency, the $O(N^3)$ scaling associated with standard GPs remains a fundamental limitation, particularly when dealing with large datasets. This scaling currently restricts the practical application of our method to individual transit windows within longer light curves, in order to keep computing time within reasonable bounds. To enable the modelling of long-duration light curves, for example those expected from the two-year single-field pointings of PLATO, more scalable techniques will need to be included in the structure learning methodology. Thankfully, there are several promising avenues for this. Hardware acceleration has been shown to enable scalable GP regression. \citet{Wang2019}, for example, have scaled exact GPs to beyond one million data points using multi-GPU parallelisation. The JAX framework underpinning \gallifrey inherently supports hardware acceleration via GPUs, and other GP implementations within the JAX ecosystem, such as \texttt{tinygp} \citep{Foreman-Mackey2024}, or the \texttt{GPyTorch} \citep{gardner2018gpytorch} in the PyTorch ecosystem, have demonstrated order-of-magnitude speed increases through GPU utilisation. Beyond hardware acceleration, exact GPs that use quasi-separable kernels, such as those introduced by \cite{Foreman-Mackey2017a}, scale linearly with data size $O(N)$ and have been shown to be a good model for stellar variability. This category of kernels is, in principle, compatible with the kernel structure learning techniques implemented in \gallifrey.  Integrating such kernels in the future would therefore potentially allow for first-principles, scalable GP structure learning for stellar background modelling. Beyond classical, exact GPs, approximate techniques such as sparse GPs \citep{Leibfried2022} and Hilbert space GPs (\citealt{Riutort-Mayol2022}), or more recently deep kernel learning (\citealt{Wilson2015, Ober2021}), hold potential for even more scalable and adaptive applications.

In conclusion, GP structure learning presents a valuable avenue for exoplanet research, particularly in the context of transit modelling and retrieval. By providing a user-friendly Python implementation of state-of-the-art structure learning algorithms in \gallifrey, we hope to make this methodology more accessible to researchers and help tackle the challenges of increasingly complex exoplanet datasets.

\section*{Code and data availability}
The full source code and all data used in this work are available at \url{https://github.com/ChrisBoettner/gallifrey}. An in-depth documentation, as well as tutorials to recreate all figures in this work, can be found at \url{https://chrisboettner.github.io/gallifrey}. The spectroscopic light curves for the HATS-46 b spectrum can be found at \url{https://cdsarc.cds.unistra.fr/viz-bin/cat/J/MNRAS/521/5636}.

\begin{acknowledgements}
Chris Boettner thanks the Young Academy Groningen for their generous support through an interdisciplinary PhD fellowship. He also wants to thank his supervisor, Pratika Dayal, for kind and continued support, and further wants to thank Evie Ahrer for the valuable discussions and providing him light curve data to play around with. 

The gallifrey package is a Python re-implementation of the Julia package \texttt{AutoGP.jl} \citep{Saad2023} and would have not been possible without it. Large parts of the implementation details are inspired by the fantastic packages \texttt{GPJax} \citep{Pinder2022} and\texttt{tinygp} \citep{Foreman-Mackey2024}. Data was retrieved using the via \texttt{astroquery} \citep{Ginsburg2019}, \texttt{lightkurve} \citep{2018ascl.soft12013L} and \texttt{PSLS} \citep{Samadi2019a}. Visualisations were made using \texttt{pandas} \citep{Thepandasdevelopmentteam2023}, \texttt{matplotlib} \citep{ThomasACaswell2023} and \texttt{seaborn} \citep{Waskom2021}.
\end{acknowledgements}

\bibliographystyle{aa} 
\bibliography{references.bib} 

\appendix
\section{Kernel functions and hyperparameters}
\label{sec:kernels}
The choice of the kernel function $k(x, x')$ is crucial in GP regression, as it defines the space of possible functions considered by the prior.

To produce valid covariance matrices, as required for Eqs. \ref{eq:gpsampledistribution} and \ref{eq:gppredictivedistribution}, the kernel function must conform to certain properties. Specifically, a valid kernel needs to be symmetric, i.e. $k(x, x') = k(x', x)$, and the kernel matrix $K$ should be positive semi-definite, fulfilling $\bm{v}^{\top} K \bm{v} \geq 0$ for any vector $\bm{v}$. Furthermore, kernels are called stationary if $k(x, x') = k(x - x')$, and isotropic if $k(x, x') = k(|x - x'|)$.

While any function fulfilling these conditions can serve as a kernel, there are several commonly used ones. The squared exponential or RBF kernel is often used to model smoothly varying stationary functions, whereas kernels from the Matérn family are used to model more rugged functions. Various periodic kernels exist to model periodic functions, and non-stationarity can be introduced using a linear kernel. We have compiled a list of commonly used kernels and their properties in \cref{tab:gpkernels}. All the listed kernels are implemented as atomic kernels in \gallifrey, and can be used for structure learning. Importantly, any kernel can be scaled by some variance $\sigma^2$, and kernels can be combined arbitrarily to create new, more complex kernels.

All kernels $k(x, x' ; \bm{\eta})$ in \cref{tab:gpkernels} depend on hyperparameters $\bm{\eta}$, governing their specific properties. Common hyperparameters include a length scale $l$, which controls the smoothness of functions within the function space, and the period $p$ for periodic kernels. In typical GP regression, the standard approach is to specify a kernel structure by choosing a combination of base kernels and then optimise the hyperparameters based on the dataset. Two common approaches for tuning hyperparameters are maximising the marginal likelihood $P(\mathcal{D} | k(\bm{\eta}))$, as given by Eq. \ref{eq:gpmarginalloglikelihood} (where the dependence on the kernel and its hyperparameters is made explicit), or maximising the leave-one-out cross-validation (LOO-CV) predictive probability:
\begin{equation}
    \log P_\mathrm{LOOCV} (\mathcal{D} | k(\bm{\eta})) = \sum_i^N \log P\left(y_i | \bm{x}_i, \bm{x}_{-i}, \bm{y}_{-i}; k(\bm{\eta}) \right),
    \label{eq:gploocvprobability}
\end{equation}
where $\{\bm{x}_{-i}, \bm{y}_{-i}\}$ represents the dataset with the $i$-th entry removed, and $P\left(y_i | \bm{x}_i, \bm{x}_{-i}, \bm{y}_{-i} \right)$ is given by Eq. \ref{eq:gppredictivedistribution}. Numerical studies suggest that using the LOO-CV predictive probability leads to more accurate predictions in cases of mis-specified covariance functions (i.e. kernel structures), while the marginal likelihood is optimal when the kernel structure is well specified \citep{Bachoc2013}.

\begin{table*}[ht]
    \centering
    \caption{Common kernels in GP regression.}
    \begin{tabularx}{\textwidth}{L|LRLr}
    Name & Equation & \# of Parameters & Parameter & Stationary \\
    \addlinespace
    \hline\hline
    \addlinespace
    Periodic & $\sigma^2 \exp\left(-\frac{\sin^2(\pi|x-x'|/p)}{2l^2}\right)$ & 3 & \begin{tabular}[c]{@{}l@{}}lengthscale : $l$ \\ variance : $\sigma^2$ \\ period : $p$\end{tabular} & Yes \\
    \addlinespace
    \hline
    \addlinespace
    Powered Exponential & $\sigma^2 \exp\left(-\left(\frac{|x-x'|}{l}\right)^\kappa\right)$ & 3 & \begin{tabular}[c]{@{}l@{}}lengthscale : $l$ \\ variance : $\sigma^2$ \\ power : $\kappa$\end{tabular} & Yes \\
    \addlinespace
    \hline
    \addlinespace
    \multicolumn{1}{L|}{Squared Exponential / \newline Radial Basis Function} & $\sigma^2 \exp\left(-\frac{(x-x')^2}{2l^2}\right)$ & 2 & \begin{tabular}[c]{@{}l@{}}lengthscale : $l$ \\ variance : $\sigma^2$\end{tabular} & Yes \\
    \addlinespace
    \hline
    \addlinespace
    \multicolumn{1}{L|}{Matérn $\left(\nu=\frac{1}{2}\right)$ / \newline Ornstein-Uhlenbeck} & $\sigma^2 \exp\left(-\frac{|x-x'|}{l}\right)$ & 2 & \begin{tabular}[c]{@{}l@{}}lengthscale : $l$ \\ variance : $\sigma^2$\end{tabular} & Yes \\
    \addlinespace
    \hline
    \addlinespace
    Matérn $\left(\nu=\frac{3}{2}\right) $ & $\sigma^2 \left(1+\frac{\sqrt{3}|x-x'|}{l}\right)\exp\left(-\frac{\sqrt{3}|x-x'|}{l}\right)$ & 2 & \begin{tabular}[c]{@{}l@{}}lengthscale : $l$ \\ variance : $\sigma^2$\end{tabular} & Yes \\
    \addlinespace
    \hline
    \addlinespace
    Matérn $\left(\nu=\frac{5}{2}\right)$ & $\sigma^2 \left(1+\frac{\sqrt{5}|x-x'|}{l}+\frac{5(x-x')^2}{3l^2}\right)\exp\left(-\frac{\sqrt{5}|x-x'|}{l}\right)$ & 2 & \begin{tabular}[c]{@{}l@{}}lengthscale : $l$ \\ variance : $\sigma^2$\end{tabular} & Yes \\
    \addlinespace
    \hline
    \addlinespace
    Rational Quadratic & $\sigma^2 \left(1+\frac{(x-x')^2}{2\alpha l^2}\right)^{-\alpha}$ & 3 & \begin{tabular}[c]{@{}l@{}}lengthscale : $l$ \\ variance : $\sigma^2$ \\ shape : $\alpha$\end{tabular} & Yes \\
    \addlinespace
    \hline
    \addlinespace
    White & $\sigma^2 \delta(x - x')$ & 1 & variance : $\sigma^2$ & Yes \\
    \addlinespace
    \hline
    \addlinespace
    Linear & $\sigma^2 (x \cdot x')$ & 1 & variance : $\sigma^2$ & No \\
    \addlinespace
    \hline
    \addlinespace
    Linear with Shift & $b + \sigma^2 ((x - c) \cdot (x' - c))$ & 3 & \begin{tabular}[c]{@{}l@{}}bias : $b$ \\ variance : $\sigma^2$ \\ shift : $c$\end{tabular} & No \\
    \addlinespace
    \hline
    \end{tabularx}
    \tablefoot{Each of the listed kernels is implemented as an atomic kernel in \gallifrey.}
    \label{tab:gpkernels}
\end{table*}

\section{Assessing the quality of the SMC posterior approximation}
\label{sec:smc_assessment}

The SMC algorithm implemented in \gallifrey approximates the posterior distribution over both kernel structures $k$ and their hyperparameters $\bm{\eta}$, as defined in Eq. \ref{eq:gpposteriorwithkernelprior}. Unlike MCMC, which targets convergence to a stationary distribution, SMC generates a sequence of weighted particle ensembles approximating a sequence of distributions $\{\pi_t\}_{t=1}^T$, culminating in a final weighted sample $\{ (k_i, \bm{\eta}_i, w_T^{(i)}) \}_{i=1}^N$ that approximates the target posterior $\pi_T \approx P(f | \mathcal{D})$. Assessing whether this final sample is a `good' approximation is important for reliable inference. Key diagnostics include:

\begin{enumerate}
    \item \textbf{Final weight distribution, effective sample size, and resampling history:} The ESS, given by Eq. \ref{eq:ESS}, is a central diagnostic for SMC performance. It quantifies the degeneracy of the final particle weights $w_T^{(i)}$. A low ESS indicates that the approximation is effectively supported by only a few particles, likely indicating a poor representation of the true posterior. While resampling during the SMC run helps mitigate degeneracy, the distribution of the final weights and the final ESS should still be examined. A low final ESS suggests the final approximation is unreliable or unstable. Furthermore, \gallifrey returns the resampling history after the SMC procedure. If particle degeneracy is no major issue, one would expect several rejuvenation steps between resampling steps. If the ESS is consistently low (triggering frequent resampling), increasing the number of particles ($N$), adjusting the annealing schedule, or increasing the number of rejuvenation steps may be necessary.  

    \item \textbf{Validation on unseen data:} In the transit fitting examples (Sects. \ref{subsec:transit_fit} and \ref{subsec:transmission_spectroscopy}), we masked the transit interval and used \gallifrey for interpolation, modelling the background signal as if no transit occurred. If the length of the time series allows, it may be advisable to mask additional intervals (without transits), and evaluate the quality of the approximation on these intervals where the ground truth is known.

    \item \textbf{Stability across runs:} If computational cost and time allows, a fundamental check could be running the \gallifrey SMC procedure multiple times with different random seeds but identical settings. Stable results across different runs suggests a good approximation for the posterior.

    \item \textbf{Number of particles:} In general, a larger number of particles $N$ leads to a better approximation but increases computational cost. The `correct' $N$ is problem-dependent and must usually be determined empirically. In \gallifrey, we used JAX's \texttt{pmap} function for efficient parallelisation. We therefore suggest using as many particles as there are physical devices (e.g. number of CPU cores) available, if possible. However, for the examples in Sect. \ref{sec:examples}, we find that as few as six particles yield stable results.
    
\end{enumerate}

\end{document}